\documentclass[preprint2]{proto}
\usepackage{times,amssymb}

\newcommand{\refs}{\par\noindent\hangindent=1pc\hangafter=1}
\begin{document}

\title{\textbf{\LARGE From Protoplanets to Protolife:\\ The Emergence
and Maintenance of Life}}

\author {\textbf{\large Eric Gaidos}}
\affil{\small\em University of Hawaii, USA}

\author {\textbf{\large Franck Selsis}}
\affil{\small\em Centre de Recherche Astronomique de Lyon, France}

\begin{abstract}
\baselineskip = 11pt
\leftskip=0.65in
\rightskip=0.65in
\parindent=1pc 

{\small Despite great advances in our understanding of
the formation of the Solar System, the evolution of the Earth, and the
chemical basis for life, we are not much closer than the ancient
Greeks to an answer of whether life has arisen and persisted on any
other planet.  The origin of life as a planetary phenomenon will
probably resist successful explanation as long as we lack an early
record of its evolution and additional examples.  Plausible but
meagerly-investigated scenarios for the origin of important prebiotic
molecules and their polymers on the Earth involving atmospheric
chemistry, meteorites, deep-sea hot springs, and tidal flat sediments
have been developed.  Our view of the diversity of extant life, from
which properties of a last universal common ancestor (LUCA) can be
inferred, has also improved in scope and resolution.  It is widely
thought that the geologic record shows that life emerged quickly after
the end of prolonged bombardment of the Earth.  New data and
simulations contradict that view and suggest that more than half a
billion years of unrecorded Earth history may have elapsed between the
origin of life and LUCA.  The impact-driven exchange of material
between the inner planets may have allowed earliest life to be more
cosmopolitan.  Indeed, terrestrial life may not have originated on the
Earth, or even on any planet.  Smaller bodies, e.g. the parent bodies
of primitive meteorites, in which carbon molecules and catalytic
transition metals were abundant, and in which hydrothermal circulation
persisted for millions of years, offer alternative environments for
the origin of life in our Solar System.  However, only planet-sized
bodies offer the stable physiochemical conditions necessary for the
persistence of life.  The search for past or present life on Mars is
an obvious path to greater enlightenment.  The absence of intense
geologic activity on Mars, which contributes to its inhospitable state
today, has also preserved its ancient history.  If life did emerge on
Mars or was transferred from Earth, the lack of sterilizing impacts
(due to a low gravity and no oceans) means that a more diverse biota
may have thrived than is represented by extant life on Earth.  On the
other hand, a habitable but still lifeless early Mars is strong
evidence against efficient transfer of life between planets.  The
subsurface oceans of some icy satellites of the outer planets
represent the best locales to search for an independent origin of life
in the Solar System because of the high dynamical barriers for
transfer, intense radiation at their surfaces, and thick ice crusts.
These also present equally formidable barriers to our technology.  The
``ultimate'' answer to the abundance of life in the Cosmos will remain
the domain of speculation until we develop observatories capable of
detecting habitable planets - and signs of life - around the nearest
million or so stars.\\ \\ \\}

\end{abstract}  

\section{\textbf{INTRODUCTION}}

This contribution's place as the last chapter in {\it Protostars and
Planets V} may betray a subtle conceit in how we view our place in a
cosmic order that runs from the interstellar medium to planetary
bodies.  (Read in reverse order, the chapters would suggest a more
humble search for our origins among wisps of interstellar gas and
dust.)  Nevertheless, this sequence makes sense, both in a temporal
and also a physical order: It describes a gradation in phenomena in
which physical and chemical inevitability (the laws of gravity,
classical and quantum mechanics, and electromagnetism) which govern
the collapse of the interstellar medium and the formation of stars,
are replaced by more stochastic processes such as accretion during
planet formation and evolution.  For example, it may be inevitable
that a cooling molecular cloud collapses, a disk forms, and that
runaway growth of planetesimals occurs in that disk, but the final
masses, orbits, and surface environments of planets may not be
predictable in more than a statistical sense.  Ultimately it is no
longer sufficient to describe what could happen, one must also
describe what {\it did} happen.  Whereas stars can be described by a
relatively small number of variables (age, rotation, and metallicity,
for example), planets, particularly terrestrial planets, cannot.  In
that context, the origin and survival of life might be the ultimate
contingency.

On the other hand, what little we know about the origin of life seems
to suggests some element of inevitability.  The primary constituents
of life (C, H, N, and O) are four of the five most abundant elements
in the universe.  Some of the monomeric molecules of life (amino
acids, sugars, etc.) are found everywhere. Laboratory experiments have
suggested possible pathways along which those monomers might become
polymers, make copies of themselves and interact in complex ways on
which Darwinian selection can act.  Evidence for life appears in
Earth's rock record as soon as there is any geologic record at all.

The dichotomy between chemical inevitability and historical
contingency infuses studies of the origin and propensity of life in
the universe (not to mention the question of what life is), and it has
spawned numerous popular books on the subject.  We leave resolution of
that problem to scientists-{\it cum}-philosophers.  In this review we
concentrate on those lines of inquiry that have experienced especially
fruitful development since the review of this subject by {\it Chyba et
al.} (2000) for {\it Protostars and Planets IV}, including new age
constraints on the appearance of clement environments and life on the
Earth, a re-assessment of predictions for the chemistry of the
prebiotic atmosphere and oceans, the formulation of a dynamical
scenario for a ``late'' cataclysmic bombardment that may have
profoundly influenced the emergence of life, and the development of
new theories for the origin of Earth's water.  Because science knows
so little about the origin of life on Earth and the potential
environments for its origin elsewhere, we feel it is important to be
open-minded - and even provocative - in the scenarios that we
consider.  Our review is structured as follows: We consider the timing
and environment of the origin of terrestrial life (\S 2) and our
understanding of the combination of factors that permit Earth-like
life to persist on a planet for an astronomically interesting period
of time (\S 3).  Finally, we address how the search for life elsewhere
in the Solar System and particularly for life-bearing planets around
other stars promises to ultimately inform us about the evolution of our
own habitable planet and the possibility of other origins elsewhere in
the Cosmos (\S 4).  Other relevant reviews since that of {\it Chyba et
al.} (2000) include {\it Shock et al.} (2000), {\it Kasting and
Catling} (2003), {\it Gaidos et al.} (2005) and {\it Chyba and Hand}
(2005).

\bigskip
\section{\textbf{WHEN AND WHERE DID LIFE EMERGE?}}

\subsection{\textbf{Origin of a Theory of Origin}}

Recorded speculation on the setting of the origin of life goes back at
least to ancient Greek civilization.  Thales of Miletus (640-546 BCE)
presciently suggested that all life, including humans, arose from the
single ``element'' water -- i.e. the sea.  His student Anaximander
(611-545 BCE) slightly modified his master's idea, substituting mud
for water and thus proposing the first primordial ``soup'' hypothesis.
Empedocles (490-435 BCE) further elaborated (or obfuscated) the
theory, proposing that life emerged in a random fashion from a
combination of the four classic Greek ``elements''.  The concept of
``spontaneous generation'' of life from non-living matter relied on
unsupported anecdote and uncontrolled experiment for two full
millenia, but was doomed by the invention of the compound
microscope ca. 1590, the discovery of ubiquitous microorganism by
Antonie von Leeuwenhoek a century later, and the {\it coup de grace}
delivered by Louis Pasteur's irrefutable 1864 demonstration of
microbial contamination in all previous origin-of-life experiments.
Modern inquiry into the origin of life began once science had
discovered aspects of the chemical basis for life, described the
theory of evolution by natural selection, and appreciated the age of
the Earth: In the 1920s Oparin and Haldane independently described a
new theory in which life emerged from ``prebiotic'' chemistry driven
by electricity or solar ultraviolet radiation in a reducing atmosphere
of the early Earth.

By necessity, tests of such theories have been limited to
demonstrations of plausibility by laboratory experiments.  This is
because the same geologic activity (volcanism, plate tectonics, and
metamorphism) that sustains geochemical cycles and life on Earth today
has destroyed nearly all of the earliest record of surface conditions
and possible life that could be used to test such theories.  The Earth
formed 4.56 billion years ago (Ga) but the paltry record of the first
500 million years (Myr) consists of a handful of zircon crystals as
old as 4.4 Ga ({\it Wilde et al.}, 2001) and a single outcrop of
heavily metamorphosed gneiss dated at 4.0 Ga ({\it Bowring and
Williams}, 1999).  The oldest putative evidence for life on Earth is
isotopically fractionated carbon in 3.85 Ga rocks from Akila Island
and the 3.7-3.8 Ga Isua formation in Greenland ({\it Schidlowski},
1988; {\it Mozjsis et al.}, 1996; {\it Rosing}, 1999).  However, this
evidence has recently been challenged ({\it van Zuilen et al., 2002};
{\it Fedo and Whitehouse}, 2002; {\it Mojzsis et al.}, 2002).
Likewise, the origin and provenance of the oldest (3.46 Ga) putative
microfossils, from the Apex chert in Australia ({\it Schopf and
Packer}, 1987), have been disputed ({\it Brazier et al.}, 2002; {\it
Brazier et al.}, 2004).  The biological nature of even the 3.4-3.5 Ga
fossil stromatolites, laminated microbial mats, in Australia and South
Africa ({\it Walter et al.}, 1980) has been questioned ({\it
Grotzinger and Rothman}, 1996).  Despite the controversy, it seems
likely that at least some of the evidence for life by 3.5 Ga will
withstand scrutiny and new kinds of evidence may emerge ({\it Furnes
et al.}, 2004).  However, the geologic record of the origin and
evolution of earlier, more primitive life seems irretrievably lost.

Any successful theory of biogenesis must provide a prebiotic source of
the organic monomers (e.g., amino acids and nucleotides) as a starting
point, and one or more mechanisms of chemical condensation of these
monomers into polymers and more complex molecules.  The Oparin-Haldane
conjecture of an atmospheric source assumed a reducing primordial
atmosphere containing abundant CH$_4$, NH$_3$ and H$_2$.  This
mechanism was brilliantly supported by Stanley Miller's experiment
({\it Miller}, 1953).  However, this scenario fell into disfavor upon
the development of models predicting planetary core formation was
contemporaneous with homogeneous accretion ({\it Stevenson}, 1980),
leaving the mantle depleted of metallic iron, and volcanic gases
relatively oxidized (N$_2$, CO$_2$, and H$_2$O).  Discharge
experiments with such gas mixtures fail to produce significant
quantities of organic molecules and underscore the particular
importance of CH$_4$ or H$_2$ ({\it Miller and Schlesinger}, 1983;
{\it Sleep et al}. 2004).

New models of Earth's earliest atmosphere predict chemically
significant concentrations of H$_2$ and CH$_4$: Although most of the
iron in the Earth would have been sequestered into the core, degassing
during impact of material with a carbonaceous chondrite composition
would have created a reducing atmosphere composed of CH$_4$, N$_2$,
NH$_3$, H$_2$, and H$_2$O ({\it Schaefer and Fegley}, 2005).  The
isotopic and elemental abundances of rare gases suggest that this
primordial atmosphere was lost: Massive hydrogen escape was probably
complete by 4.47~Ga ({\it Podesek and Ozima}, 2000) and the atmosphere
was closed to all elements except H and He by 4.3 Ga ({\it Tolstikhin
and O'Nions}, 1994).  However, this still leaves a period of between
30 and 200~Myr after core formation in which a Urey-Miller atmosphere
could have existed, perhaps plenty of time for biogenesis to occur.
Furthermore, hydrogen out-gassing later from volcanoes may have been
more strongly retained by an anoxic atmosphere where the upper
atmosphere did not contain singlet oxygen that absorbs extreme
ultraviolet radiation from the Sun ({\it Tian et al.}, 2005), although
this is not conclusive ({\it Catling}, 2006).  Nevertheless,
alternative sources of organic monomers are available: One appeared
serendipitously in the form of a meteorite which fell near the town of
Murchison, Australia, in 1969.  The archetype CM meteorite was found
to contain a suite of organic molecules including many of the biotic
amino acids (see review by {\it Ehrenfreund et al.} 2001).  Both
meteorites and comets might have provided organics to the early Earth
({\it Chyba et al.}, 1990).

A decade after the Murchison meteorite fell, the first deep-sea hot
spring chemotrophic ecosystem supported by the mixing of sulfidic
hydrothermal fluids with oxygenated seawater was discovered ({\it
Corliss et al.}, 1979).  The appreciation that microorganisms could
have colonized such high-temperature settings and exploited chemical
energy sources before the advent of photosynthesis led to interest in
their potential role in the origin and early evolution of life.
Currently, the hypothesis of a hydrothermal origin of life draws
support from three observations: First, hydrothermal systems are sites
where organic synthesis is thermodynamically favored ({\it Shock and
Schulte}, 1988; {\it Shock et al.}, 2001).  Second, these environments
contain abundant iron and nickel sulfides that may catalyze reactions
of potential prebiotic importance ({\it Huber and
W\"{a}chtersh\"{a}user}, 1997) and are present as co-factors in many
enzymes ({\it Johnson et al.}, 2005).  {\it Cody et al.}  (2000)
showed that reaction of iron sulfide (FeS) with alkyl thiols (RSH),
where R is an alkane group, produces carbonylated iron-sulfur compounds via
\begin{equation}
{\rm 2FeS + 6CO + 2RSH \rightarrow Fe_2(RS)_2(CO)_6 + 2S^0 + H_2},
\end{equation}
which they suggest to be responsible for catalysis, in lieu of mineral
surfaces themselves.  (The possible role of metal sulfides in
prebiotic chemistry and subsequent incorporation into central
metabolic pathways has been recently reviewed by {\it Cody} (2004).
{\it Holm and Andersson} (2005) discuss the challenges of conducting
hydrothermal chemistry under geologically relevant conditions.)
Third, many thermophilic and hyperthermophilic archaea and bacteria
are located near the root of phylogenetic trees constructed from small
subunit ribosomal RNA gene sequences.  This has been taken to suggest
that a primitive character of the last universal common ancestor of
all life was adaptation to high temperature, as originally suggested
by {\it Woese} (1987), an inference widely, but not completely,
accepted ({\it Galtier et al.}, 1999; {\it Brochier and Philippe},
2002; {\it Di Giulio}, 2003).  (See the next section for an
alternative explanation of thermophily.)

Another successful conjecture in origin of life studies is the idea of
an ``RNA world'' in which ribonucleic acid (RNA) played the role of
both DNA and protein in primitive organisms by carrying information
and catalyzing chemistry ({\it Orgel}, 1968; {\it Crick}, 1968; {\it
Gilbert}, 1986).  This conjecture is supported by the appearance of
RNA in ubiquitous and highly conserved - and thus evolutionarily
ancient - parts of the cellular machinery such as the ribosome, the
demonstration that ribonucleotides are catalytically active ({\it
Cech}, 1986), and by the success of evolving catalytically small,
active RNA molecules in the laboratory ({\it Joyce}, 2004).  In
contrast to the hypothetical high-temperature origin of life described
above, the phosphodiester backbone of RNA and the nucleobases
themselves are unstable under high-temperature aqueous conditions
(e.g., {\it Levy and Miller}, 1998).  One scenario is for an RNA world
to evolve under near-freezing conditions, perhaps in pockets of
eutectic brine within ice where components were cyclically frozen and
re-hydrated ({\it Orgel}, 2004; {\it Vlassov et al.}, 2005).  Brines
have also been suggested as the site of prebiotic purine and
pyrimidine synthesis and polymerization ({\it Bada et al.}, 1994, {\it
Miyakawa et al.}, 2002a, {\it Miyakawa et al.}, 2002b).

Recently, investigators have turned to wet-dry cyclic chemistry at
clement temperatures, perhaps driven in the sediments of intertidal
flats. {\it Commeyras et al.}  (2001) describe a mechanism of
prebiotic polypeptide synthesis through cyclic condensation with
N-carbonyl amino acids under alternating pH conditions in the presence
of significant nitrogen oxides in the atmosphere.  [See also {\it
Lathe} 2004 for a speculative scenario based on salt concentrations.]
Alternatively, a more stable predecessor to RNA such as a peptidal
molecule has been posited.  {\it Russell and Arndt} (2005) argue for
biogenesis at low-temperature, alkaline submarine seeps.  These seeps
form mounds containing precipitated iron-nickel sulfides through which
strong chemical gradients are maintained between the H$_2$-rich,
reducing fluids and more oxidizing oceans and driving the reduction of
CO$_2$ or HCO$_3^-$ to acetate (COOH).

If core metabolism reflects a hydrothermal environment, and RNA
evolved before protein, then the thermal instability of RNA suggests
that it in turn was preceded by an unknown protobiotic world that
functioned at higher temperatures, and therefore the thermophilic
character of a last universal common ancestor is unrelated to a high
temperature origin of metabolism.  Alternatively, RNA and the core
metabolism of extant organisms appeared in different lineages.  These
considerations suggest a substantial evolutionary history preceding
LUCA.  Such a history may have involved the extensive chimerism of
lineages that evolved from different environments.  An analogous
history is recorded in the complex organelle structure of eukaryotic
microalgae that have experienced engulfment and endosymbiosis of
independent unicellular lineages ({\it McFadden}, 2001).  {\it Woese} (2000) has
suggested that the earliest history of RNA/DNA-based life was marked
by the rampant ``horizontal'' transfer of genes between organisms,
absence of distinct lineages and communal evolution of the gene pool.
Less efficient and redundant components could have been discarded
(e.g., the information-carrying molecules in the high temperature
contributor, the metabolic machinery in the low-temperature
contributor), leaving an organism whose chemical ancestry derives from
very different environments.

Furthermore, the environment(s) in which the origin of life took place
need not resemble any environment on the modern Earth, and indeed may
not be habitable by the standards of modern organisms.  The evolution
of life may have involved ``frozen accidents'' in which universal
biological attributes selected for in an archaic environment are
retained, even in the face of maladaptation in a new environment,
because any changes in them would be too costly to the fitness of
organisms.  For example, while the eukaryotic cell may have arisen
from a chimeric fusion of representatives of the Bacteria and Archaea,
both domains of life that contain species that thrive at temperatures
near 100$^{\circ}$C, no eukaryote has been found that grows at
temperatures above $\sim 60^{\circ}$, probably because the
incorporation of membrane-surrounded organelles such as the nucleus
requires permeability that renders the membrane susceptible to
destruction at high temperatures.  It is conceivable that life arose
at temperature exceeding 120$^{\circ}$C, but that the universal use of
lipid membranes for structure and triphosphates for energy has
rendered those environments forever inaccessible to life.

Darwin's proposal that all life on Earth shares a common ancestry is
supported by vast amounts of molecular work.  Yet, much of the
microscopic world is classified only by molecular techniques such as
the polymerase chain reaction (PCR) and it is conceivable that
completely ``alien'' organisms based on different molecules flourish
undetected under our feet ({\it Davies and Lineweaver}, 2005).  If all
Earthly life does have a single origin this might mean that the origin
of life is sufficiently infrequent that the probability of it
happening more than once on the same planet is low.  Alternatively, it
might mean that sometime in Earth history all other forms of life went
extinct.  Although it may be chauvinistically satisfying to think that
other forms of life were out-competed by our common ancestor, Nature
tolerates the competitive or non-competitive co-existence of countless
forms of life, often within the same ecological niche (e.g., there are
300,000 known species of plants).  Although there is no evidence that
independent forms of life ever existed, it is difficult to exclude
them from the first billion years of history in the absence of
morphological fossils, and impossible to exclude them from the first
600 Myr as there is no record at all.  Such a loss in diversity would
not be the first to be inferred in the history of life.  For example,
the diversity of animal body plans recorded in fossil deposits of
exceptional preservation such as the Burgess Shale is thought to
greatly exceed later bodyplan diversity.  Instead of competition,
perhaps a uniquely catastrophic event extinguished all but a few,
related forms of life that occupied some refuge.

\subsection{\textbf{Impacts, Bottlenecks, and Frozen Accidents}}

Giant impacts capable of vaporizing the oceans may have provided such
an extinction event.  A ``late'' (3.9 Ga) episode of impacts is
recorded on the Moon and in the Martian meteorite ALH 84001 ({\it
Turner et al.}, 1997).  Sterilizing impacts may have limited the
emergence of life ({\it Maher and Stevenson}, 1988) and imposed a
high-temperature ``bottleneck'' through which only adapted organisms
could have passed, thus explaining the inference that LUCA was a
thermophile ({\it Sleep and Zahnle}, 1998, {\it Nisbet and Sleep},
2001).  Giant impacts may also have contributed to the destruction of
the rock record itself.  One model for this ``late heavy bombardment''
involves the decay of a long-lived reservoir of impactors somewhere in
the outer Solar System ({\it Fernandez and Ip}, 1983).  However,
searches for geochemical evidence for an extraterrestrial input to the
Earth system at 3.8-3.7 Ga have yielded ambiguous results ({\it Anbar
et al.}, 2001; {\it Schoenberg et al.}, 2002; {\it Frei and Rosing},
2005).  A null result from such searches supports an alternative
scenario in which the impacts occurred in a single cataclysm ca. 3.9
Ga ({\it Dalrymple and Ryder} 1993; {\it Cohen et al.} 2000).  Such an
event can be produced by a 1:2 mean motion resonance crossing of
Jupiter and Saturn ({\it Tsiganis et al.}, 2005) during an early
period of giant planet migration driven by planetesimal scattering
({\it Hahn and Malhotra}, 1999).  This scenario is consistent with
evidence for an asteroidal origin of the impacts ({\it Strom et al.},
2005).

Previously, the earliest, evidence for life in the rock record,
apparently at the tail end of a continuous period of sterilizing giant
impacts, was taken to suggest that the origin of life was geologically
instantaneous and would occur just as quickly on other planets were
conditions correct (e.g. {\it Lineweaver and Davis}, 2002).  If the
scenario of a 'brief' cataclysm is correct, life may have emerged
during the previous 600 Myr period that followed a magma ocean ({\it
Boyet and Carlson}, 2005) and Moon-forming impact ({\it Lee et al.},
2002) at around 4.5 Ga.  During that time the impact rate may have
been permissible for life, and considerable prebiotic and biological
evolution could have taken place of which we have no record.  Or do
we?  Assuming that life emerged prior to 3.9 Ga and survived the
impact bottleneck in deep refugia, the genetic information carried in
the last universal common ancestor(s) might tell us something about
that early environment.  For example, oxygen in a pre-3.9 Ga
atmosphere would explain the paradox of the presence of cytochrome
{\it c} terminal oxidases in many species of both bacteria and
archaea, and thus presumably in a LUCA, and before the origin of
oxygenic photosynthetic cyanobacteria ({\it Castersana et al.}, 1994).
A giant ocean-vaporizing impact would extinguish photosynthetic life,
but perhaps not deeper-living organisms that had profited from that
oxygen (such as those that exist in modern vent systems).  A narrow
bottleneck would be a convenient explanation for why only one form of
life exists on modern Earth.  The requirement of giant planets near a
resonance suggests that such cataclysms may not occur (or may occur at
a different time) in extrasolar planetary systems with different giant
planet architectures.

Impacts also provide a mechanism by which life might be transferred
from one planet to another.  Interest in the interplanetary transfer
of life (related to, but to be distinguished from to conjectures of
cosmological ``panspermia'') was catalyzed by the discovery of
meteorites from Mars, the elaboration of the spallation mechanism of
impact ejection ({\it Melosh}, 1984), and dynamical simulations
showing small but finite probabilities that such ejecta could be
transferred between the inner planets on timescales of thousands of
years or less ({\it Gladman and Burns}, 1996).  Magnetic constraints
on the thermal history of the ALH~84001 meteorite during the $\sim$17
Myr transit ({\it Goswami et al.}, 1997) are permissive of life ({\it
Weiss et al.}, 2000).  Laboratory experiments indicate that bacteria
and their spores can survive the shock pressures and acceleration
associated with impact ejection ({\it Mastrapa et al.}, 2001; {\it
Burchell et al.}, 2001; {\it Burchell et al.}, 2003; {\it Burchell et
al.}, 2004) and can find sufficient protection from radiation within
rock fragments a few cm in size ({\it Horneck et al.}, 2001).

Transfer between the inner planets may have been a ubiquitous process.
Simulations by {\it Gladman et al.} (2005) show that 1\%, 0.1\% and
0.001\% of ejecta from Earth reach Earth, Venus and Mars in 30,000
years.  In the first case, this suggests that ejecta may have been a
refugia for microorganisms during a giant impact event in which
sterilizing conditions existed for thousands of years ({\it Wells et
al.}, 2003).  Alternatively, ejecta on ``express'' trajectories (a few
years) could have reseeded planets after giant impact extinction
events, provided there was a second, life-bearing planet.  Climate
models suggest that Venus, if it started out with an Earth-like
inventory of water, could have experienced clement surface
temperatures ({\it Kasting et al.}, 1993) and there is
geomorphological evidence for a very early warm, wet Mars ({\it
Jakosky and Phillips}, 2001).  Even if sterilizing impact was
inevitable on each planet, the probability of simultaneous events
(within a few thousand years) on the two planets would be vanishingly
small.  This could mean a novel requirement for planetary
habitability, that of a second habitable planet.

If life can be transferred between planets then it is not too great a
leap of logic to suppose that it arose on another planet and was later
transferred to Earth.  [Although it appears unlikely that meteorites
could be exchanged between planetary systems ({\it Melosh}, 2003; {\it
Wallis and Wickramasinghe}, 2004) it was more likely for stars
(possibly like the Sun) formed in a dense cluster ({\it Adams and
Spergel}, 2005).] Mars is {\it a priori} the favorite alternate planet
of origin because of its lower escape velocity and because there is
evidence for at least episodic Earth-like conditions in the past -
although the exact conditions are controversial ({\it Carr}, 1999;
{\it Craddock and Howeard}, 2002; {\it Bhattacharya et al.}, 2005).
There is no such evidence (one way or another) for Venus and it has a
deeper gravity well.  {\it Sleep and Zahnle} (1998) have also found
that any organisms on Mars would have been more likely to survive
giant impacts in the past, again because the kinetic energy of the
impact is smaller, and the absence of the latent heat of fusion of a
vaporized global ocean which would delay cooling (assuming Mars had no
such ocean).  However speculative such theories may seem, the absence
of any record of early life on the Earth suggests that we keep an open
mind on such matters.

\subsection{\textbf{Life first, planets second?}}

Indeed, planetary bodies much smaller than Mars represent a potential
site for the origin (but not maintenance) of Earth life.  Many
carbonaceous chondrite meteorites record geochemical alteration by
liquid water, and it is presumed that they originate from parent
bodies a few tens of km across, i.e., large enough to have maintained
temperatures above the freezing point of water for millions of years,
but too small to have experienced differentiation and high-temperature
metamorphism ({\it Keil}, 2000).  The main asteroid belt presently
contains more than 300 asteroids with diameters larger than 50~km and
the primordial belt may have contained 10$^3$-10$^4$ times as many
({\it Weidenschilling}, 1977).  A scenario for the origin of life in a
primitive planetesimal and its subsequent transfer to Earth would
involve biogenesis while liquid water was present, transfer of
protoorganisms to the Earth after the Moon-forming impact
approximately 30~Myr into Solar System history ({\it Jacobsen}, 2005),
and preservation of the organisms during any intervening period.  This
scenario is distinct from the survivability of organisms in asteroids
to the present day, which {\it Clark et al.} (1999) have dismissed
based on thermal, radiation, and energetic arguments.

Carbonaceous chondrite meteorites contain abundant (up to a few weight
percent) water.  Masses of several main belt asteroids determined by
the orbits of satellites give low densities suggestive of high water
ice content and/or high void space ({\it Marchis et al.}, 2005) and
consistent with a picture of an asteroid as an icy ``rubble pile''
({\it Weidenschilling}, 1981). Highly permeable, water-rich asteroids
would have been sites of hydrothermal circulation early in their
history.  Water in the interior of parent bodies would be liquefied
and mobilized by the heat from decaying $^{26}$Al and $^{60}$Fe while
protected by an ice-filled impermeable crust a few km thick.
Additional internal heat can be provided by serpentinization (see the
chapter by {\it Jewitt et al.}) and possibly impacts.  Detailed
three-dimensional simulations of hydrothermal convection in a 40~km
body show interior temperatures remain well above the freezing point
for millions of years ({\it Travis and Schubert}, 2005).

Carbonaceous chondrites (and by inference their parent bodies) also
contain organic molecules, including amino acids ({\it Kvenvolden et
al.}, 1970) and polyhydroxylated compounds (e.g., sugars) ({\it Cooper
et al.}, 2001), and their possible role as a source of important
biotic precursor molecules has long been scrutinized.  The stable
isotopes of C and N in this organic matter suggests an origin in the
interstellar medium ({\it Alexander et al.}, 1998), but
significant processing could have occurred in the solar nebula and in
meteorite parent bodies.  Although aqueous alteration in many parent
bodies involved relatively oxidizing conditions and thus led to loss
of organic material (e.g., conversion to CO$_2$ and carbonates) ({\it
Naraoka et al.}, 2004), a few meteorites, particularly CM meteorites
like Murchison, seemed to have been altered by reducing fluids ({\it
Browning and Bourcier}, 1996).  Moreover, {\it Shock and Schulte}
(1990) make thermodynamics arguments for amino acid synthesis by
aqueous alteration of polycyclic aromatic hydrocarbons (PAHs) a common
organic in the interstellar medium and primitive meteorites, and
Strecker synthesis by reaction of ketones or aldehydes with HCN and
NH$_3$ ({\it Shulte and Shock}, 1992).

{\it Clark et al.} (1999) argue that the emergence of endogenous
 organisms is {\it a priori} less likely in an asteroid than on a
 planet because the former are smaller, and because they supposedly
 comprise less diverse environments.  However, the macroscopic scale
 of an environment is unlikely to affect its potential to host
 microscopic prebiotic chemistry.  First-order chemical kinetics
 depends on the {\it concentration} of reactants rather than the total
 molar quantity and high concentrations of reactants (the ``soup'')
 are more plausibly produced in small environments (``puddles'') than
 in large ones.  If the first steps in the origin of life consist of
 prebiotic chemistry, it is chemical diversity rather than physical or
 geologic diversity that is important.  Melting and high-temperature
 metamorphism associated with the accretion and differentiation of
 planetary embryos and planets results in chemical equilibrium and the
 destruction of chemical diversity.  Besides many of the important
 terrestrial minerals such as olivines, pyroxenes, and clays,
 meteorites contain a diverse suite of minerals that have not been
 found on Earth, including various metal sulfides and phosphates
 (Table \ref{tab.minerals}).  Carbonaceous chondrite meteorites also
 contain abundant metallic iron-nickel grain, in contrast to the
 surface of the Earth where such metal alloys are extremely rare and
 found only associated with ophiolites (preserved pieces of oceanic
 crust that have been heavily altered by the reducing fluids
 associated with serpentinization).  As discussed above, metal
 sulfides and metals may have played an important catalytic role in
 prebiotic chemistry.

\begin{table}[ht]
\caption{Uniquely extraterrestrial minerals}
\begin{center}
\label{tab.minerals}
\begin{tabular}{ll}
\hline
Name & Chemical formula \\
\hline
barringerite & Fe$_{2-x}$Ni$_x$P \\
brezinaite & Cr$_3$S$_4$ \\
brianite & Na$_2$CaMg(PO$_4$)$_2$ \\
carlsbergite & CrN \\
daubreelite & FeCr$_2$S$_4$ \\
farringtonite & Mg$_3$(PO$_4$)$_2$ \\
gentnerite & Cu$_8$Fe$_3$Cr$_{1}$S$_{18}$ \\
haxonite & Fe$_{20}$Ni$_3$C$_6$ \\
heideite & (Fe,Cr)$_{1+x}$(Ti,Fe)$_2$S$_4$ \\
krinovite & NaMg$_2$CrSi$_3$O$_{10}$ \\
lawrencite & (Fe,Ni)Cl$_2$ \\
majorite & Mg$_3$(Fe,Al,Si)$_2$(SiO$_4$)$_3$ \\
merrihueite & (K,Na)$_2$(Fe,Mg)$_5$Si$_{12}$O$_{30}$ \\
\hline
\end{tabular}
\end{center}
\end{table}

Although these parent bodies were small, they were extremely numerous
and diverse.  Each of these bodies would have differed because of
chemical gradients in the solar nebula, their precise accretion
history, and their final size.  The simulations of {\it Travis and
Schubert} (2005) also show that within a single (undifferentiated)
body there is a diversity of hydraulic histories and presumably,
degrees of chemical alteration.  Individual impacts at speeds low
enough to be non-sterilizing would induce additional heterogeneity in
physiochemical conditions.  Essentially, each of these bodies would
represent a different ``experiment'' in low-temperature inorganic and
organic chemistry.  Many or most of these experiments would be cut
short by accretion onto large embryos where melting and
differentiation would occur.  However enough bodies might have
survived the 30 Myr during which accretion of the Earth was completed.
Disruption of these bodies by mutual collisions induced by the
gravitational perturbations of Jupiter might produce frozen fragments
containing protolife that could successfully transit the thick
atmosphere of an abiotic Earth to thaw on and colonize its surface.

Could some form of protolife have emerged in a primordial asteroid and
then persisted long enough (perhaps in a frozen state) to await
collisional disruption of the body into fragments small enough for a
relatively gentle arrival on the Earth?  Such a scenario requires that
(1) life evolved ``very quickly'' (within a few to tens of Myr); (2)
that it was preserved in the parent body or fragments of the parent
body during the period of the formation and cooling of the terrestrial
planets (perhaps 30-100 Myr), (3) that it was successfully transferred
to the Earth (or Mars) intact, perhaps in a small fragment; and (4)
that it arrived in an environment in which it could thrive.

The unsuccessful (or overly successful) search for fossil life in
meteorites has been well documented, e.g. {\it Anders et al.} (1964).
If life did emerge in the interior of primitive planetesimals, why has
it or evidence for biological activity not been found in a collected
primitive meteorite?  One possibility is that any organisms or
biomarkers have been degraded by radiation or impacts over the
intervening 4.5 billion years since these bodies were warm.
Furthermore, only the small fraction of organics that are soluble have
been thoroughly studied.  The remainder is thought to be dominated by
complex (poly)aromatic hydrocarbons ({\it Cody et al.}, 2002; {\it
Sephton et al.}, 2003).  There are controversial measurements of
L-excess chirality of meteoritic amino acids ({\it Engel and Nagy},
1982; {\it Pizzarello and Cronin}, 2000).  Another explanation is that
the world's meteorite collection probably samples only $\sim$100
parent bodies in the present asteroid belt.  Finally, the population
of bodies that could have seeded Earth within a few tens of Myr has
been completely depleted over the age of the Solar System.  In other
words, if terrestrial life did emerge in a planetesimal, then we do
not find it in our meteorites because that body or its fragments
already arrived long ago, and we, and all life on Earth, are the
result.

The scenario that life arises in the interior of undifferentiated,
primitive body and subsequently found a permanent home on a
differentiated planet requires a population of small bodies with a
dynamical lifetime longer than (but not much longer than) the
accretion time scale of a potentially habitable planet.  Terrestrial
planet formation is a relatively efficient process, i.e. most
planetesimals are accreted into large embryos (which differentiate and
melt) rather than small bodies, nevertheless final clearing may take
well over 100 Myr ({\it Goldreich et al.}, 2004).  In addition, the
gravitational perturbation of a gas giant planet such as Jupiter
inhibits planet formation and scatters bodies at large distances.
Thus, the formation of a giant planet and the equivalent of an
asteroid belt may be a prerequisite for the emergence of life in a
planetary system.

\bigskip
\section{\textbf{ELEMENTS OF HABITABILITY}}

\subsection{\textbf{The Habitable Zone}}

Once life is established on a planet, and assuming it survives
catastrophes such as giant impacts, what factors are important to its
persistence over a significant (i.e. observable) period of time?  The
range of orbital semi-major axes for which the surface temperatures on
Earth-like planets would permit liquid water describes a ``habitable
zone'' around a star ({\it Huang}, 1959).  This will change with
stellar luminosity evolution ({\it Hart}, 1979) and will depend on the
concentration of greenhouse gases in the atmosphere and therefore on
geochemical feedbacks ({\it Kasting et al.}, 1993) and rates of
geologic activity such as volcanism ({\it Franck et al.}, 2000).  That
region of space in which a planet on a stable orbit will remain in the
habitable zone over an extended period of time is known as the
continuously habitable zone.  The Earth's orbit is relatively stable
against the perturbations of the other planets over billion-year
timescales ({\it Laskar}, 1994).  It will remain in the habitable zone for
another 1-2 billion years before experiencing a runaway greenhouse
({\it Caldeira and Kasting}, 1992).

However, the known systems of extrasolar planets have giant planet
configurations quite unlike that of our Solar System.  Yet unseen
terrestrial planets in the habitable zones of these stars may have
orbits that are dynamically unstable against gravitational
perturbation by the detected giant planets.  The criterion of
dynamical habitability has motivated a host of publications that
explore the stability of small (i.e. massless) planets within known
giant planet systems ({\it \'{E}rdi et al.}, 2004; {\it Asghari et
al.}, 2004; {\it Ji et al.}, 2005; {\it Jones et al.}, 2005, see also
references in {\it Gaidos et al.}, 2005).  These show that small
planets could persist in the habitable zone of some, but not all these
systems for the duration of the simulations (which tend to be limited
to millions of years).  The kinematics of hypothetical extrasolar
planets and the implications for habitability have been less explored:
In the presence of at least two other planets, planets may experience
chaotic obliquity fluctuations.  The presence of oceans would moderate
surface temperatures, however, making them habitable at least for
simple life ({\it Williams and Pollard}, 2003).  A similar conclusion
is reached for planets on eccentric orbits ({\it Williams and
Pollard}, 2002). Planets on the close-in habitable zones around much
fainter M stars will be subject to tidal locking however even in this
case sufficient convective heat transport to the dark side can
maintain atmospheres against collapse ({\it Joshi et al.}, 1997).
Although we may have a quantitative understanding of the allowed
ranges of orbital and rotation necessary for the habitability of an
Earth {\it twin}, many other factors determine whether a planet can
support life ({\it Taylor}, 1999).  Some of these, including the
frequency of supernovae and giant impacts have been explored by {\it
Gonzalez et al.} (2000).

\subsection{\textbf{Planetary Water}}

Water is an indisputably indispensable commodity of planetary
habitability and a defining constituent of Earth's surface.  Any model
of terrestrial planet habitability must include a component that
addresses the abundance of water, and any such component must
satisfactorily explain the origin of Earth's water.  The inner regions
($\sim$1~AU) of model primordial solar nebulae are devoid of water, as
a consequence of diffusion of water vapor outwards along a thermal
gradient and condensation at a ``snow line'', and in apparent
agreement with the correlation between the water content and the
orbital distance of asteroids (assumed to be their formation
distance).  It is also thought that retention of water against loss to
space is efficient only when a planet had grown to a certain mass.
Compared to the abundance of water in primitive materials such as CI
chondrites (1-10\%), indeed the bulk Earth is dry; roughly 0.023\% by
weight for the oceans and an uncertain but probably similar amount for
the water in the hydrous mantle ({\it L\'{e}cuyer}, 1998).  Rare gas
isotopic and elemental abundances also indicate the loss of copious
hydrogen to space ({\it Pepin}, 1991) and since water is the major
reservoir of hydrogen (at least on the modern Earth) and this must be
accounted for as well (see below).

The accretion of a late ``veneer'' of water-rich material has been
postulated as the source of Earth's water.  Water-rich carbonaceous
chondrite meteorites were early suspects ({\it Boata}, 1954).
Observations and models of the solar nebula suggest that bodies beyond
2.5~AU may be water rich and the source of carbonaceous chondrites.
The relative abundance of deuterium to hydrogen of H$_2$O in these
meteorites spans the value of seawater ($1.53 \times 10^{-4}$).  (In
these discussion, it should be kept in mind that the material that was
the source of Earth's water may not have any representatives in our
meteorite collections or indeed in the Solar System; terrestrial
planet accretion is a relatively efficient process!)  A major
contribution by comets ({\it Chyba}, 1987), is not consistent with the
D/H values nor the abundances of rare gases ({\it Dauphas and Marty},
2002) and is dynamically difficult.  Another mechanism of inwards
water transport is the condensation of ice grains beyond the ``snow
lines'' where temperatures are below 160~K, inwards migration by gas
drag, and sublimation ({\it Cyr et al.}, 1998, {\it Cuzzi and Zahnle},
2003, {\it Mousis and Alibert}, 2005).

New developments in isotopic geochemistry and numerical dynamics
calculations have added substance to investigations of the source and
timescales of delivery of Earth's water.  Investigators have sought to
use the abundance of siderophilic elements (Ni, Co, Ge, and the
platinum group elements) in the Earth's crust as a constraint on any
``late'' (post core-formation) accretion of primitive material onto
the Earth ({\it Chyba et al.}, 1990).  {\it Righter} (1997) has
proposed that the high abundances are instead controlled by
equilibration with metallic iron at the base of an early magma ocean.
New results illuminate, but don't resolve, this controversy; neodynium
isotope data support the existence of a magma ocean ({\it Boyet and
Carlson}, 2005) but new high-pressure experiments for some elements
have not supported Righter's explanation for crustal siderophile
abundances ({\it Holzheid et al.}, 2000; {\it Righter}, 2003; {\it
Kegler et al.}; 2005).  Based on analysis of the hafnium-tungsten and
samarium-neodynium isotope systems, the bulk of the Earth is now
thought to have accreted in about 10 Myr, and was essentially complete
at 30 million years ({\it Jacobsen}, 2005; {\it Boyet and Carlson},
2005).  Rapid accretion of the Earth makes the delivery of
siderophilic elements more dynamically plausible since complete
clearing of planetesimals may have taken as long as 300 Myr ({\it Goldreich
et al.}, 2004).  This implies that dehydrated but undifferentiated
material near Earth's orbit supplied the siderophilic elements - but
no water.  (Of course, those same simulations fail to produce the
Earth in the required 30 Myr!)

Numerical simulations have been employed to investigate mechanisms by
which water-bearing material beyond 2.5~AU might be transported
inwards to the orbit of the Earth.  The late impactor cataclysm
scenario described in {\it Gomes et al.} (2005) is not a contender as
the event occurs long after the earliest evidence for water on the
planet, i.e. the isotopic composition of oxygen in 4.4-4.3 Ga zircons
({\it Mojzsis et al.}, 2001).  [Zircons are abundant in granitic rocks
produced by partial melting in the presence of water, but zircons have
also been found in lunar igneous rocks ({\it Meyer et al.}, 1996)].
Also, the estimated total accreted mass is too low to supply the
water.  An alternative mechanism is that self-scattering of planetary
embryos (and their water) in the late stages of planetary accretion
moved water inwards ({\it Morbidelli et al.}, 2000).  N-body
simulations ({\it Chambers and Wetherill}, 1998; {\it Chambers}, 2001)
suggest that the Earth is the result of the fusion of tens of
individual planetary embryos, which formed within a broad range of
orbital distances. Some of them may originate from regions at or
beyond 2.5 AU where hydrated minerals or even ices were stable. Only a
small number of these volatile-rich embryos are expected to contribute
to the formation of an Earth at 1~AU but a single Moon-sized embryo
formed at 3 AU and made of 10\% water by mass would give the Earth the
equivalent of 5 modern oceans. In this scenario, the delivery of water
to the telluric planets by ``wet'' embryos from more distant parts of
the primordial solar system is a stochastic process relying on a small
number of collisions.  As a consequence, the water content of
terrestrial planets is expected to be variable, even within a single
planetary system.  {\it Raymond et al.} (2004) carried out simulations
of embryo scattering and accretion terrestrial planet formation with
different nebular solid densities, position of the ``snow line'', and
orbit of an outer giant planet.  The vast majority of planets that
formed in the ``habitable zone'' (0.8-1.5 AU) had water inventories
equal to or greater than that of the Earth. They found that the
terrestrial planets in their simulations ended with an average water
abundance about that of Earth, as long as the giant planet
configuration was not too different from the one in the Solar
System. They showed that dry planets and extremely water-rich planets
can also be expected

This mechanism of water delivery can explain the difference in the
water inventories of Earth and Mars: At the orbital distance of Mars,
planetary formation is less efficient because of the influence of
Jupiter, and Mars can be a remaining dry embryo (or the result of a
very small number of dry embryos) formed locally and to which water
was only brought by the late bombardment ({\it Lunine et al.}, 2003).
However, some discrepancies between N-body simulations and
observations still need to be explained. {\it Wiechert et al.} (2001)
pointed out that the identical isotope fractionation of oxygen on the
Earth and the Moon implies a similar composition of the Moon-forming
impactor ``Theia'' and the proto-Earth.  Oxygen isotopic fractionation
is a signature of the heliocentric distance of formation.  Even if
Earth and Theia formed at the same distance from the Sun ({\it
Belbruno and Gott III}, 2005) it is difficult to explain how Theia and
the proto-Earth could have shared the same isotopic signature.
Although oxygen isotopes might have been homogenized in the
circumterrestrial disk in the aftermath of the giant impact ({\it
Pahlevan and Stevenson}, 2005) this would not explain the
terrestrial-like superchondritic $^{142}$Nd/$^{144}$Nd ({\it Boyet and
Carlson}, 2005).

Another potential issue with the delivery of water by embryos is its
escape from the embryos themselves.  ``Wet'' embryos formed from
km-sized objects in $\sim10^4$~yr but were unable to radiate away the
energy of accretion ($>3GM^{2}/5R$) in this period because the
required cooling rate exceeds (by orders of magnitude) the
$\sim$300~W~m$^{-2}$ runaway greenhouse limit. This created a ``magma
ocean'' phase, during which a dense steam atmosphere equilibrated with
a molten rocky surface ({\it Zahnle}, 1998).  For embryos with masses
between 0.01 and 0.1 Earth masses, this phase lasted 0.5 to 4~Myr,
which is comparable to the typical lifetime for protoplanetary gas
disks ({\it Lyo et al.}, 2003; {\it Armitage et al.}, 2003).  While
the disk was present, its opacity screened the embryos from intense UV
radiation from the young star ({\it Ribas et al.}, 2005).  Once the
disk is absent, however, this radiation can drive photolysis of water
in the upper atmospheres of water and escape of hydrogen to space.
Furthermore, if core formation in these embryos is incomplete, water
reacts with iron in the mantle, releasing large amounts of molecular
hydrogen ({\it Zahnle}, 1998).  Escape to space of hydrogen from the
relatively low gravitational potential of lunar-sized embryos would be
efficient.  The history of water may be very different in the inner
regions of planetary systems that hosted different-sized embryos (due
to a different mass surface density and isolation mass, for example)
or had a different disk lifetime than that of our Solar System.

How much water is ``enough'', and where does it end up?  {\it Matsui
and Abe} (1986) showed that the amount of water at Earth's surface is
roughly what would expect were it controlled by the solubility of
water in silicate melt, i.e. an early magma ocean.  Besides the
reservoir of the global ocean, a significant amount of water may be
sequestered in the mantle.  The concentration of water in Earth's
mantle is a subject of active research ({\it Tarits et al.}, 2004) but
it may be the equivalent of several oceans ({\it Litasov et al.},
2003).  A significant amount of water could have been lost as the
hydrous silicates reacted with metallic Fe during core formation to
form iron hydrides (FeH$_x$) that would be sequestered into core.  The
residual oxygen then reacted with ferrous iron in the mantle.  {\it
Hirao et al.} (2004) estimates that the core could contain H that is
the equivalent of 8-24 oceans of water.  Water may also have been lost
by erosion of the atmosphere by giant planets, and (as hydrogen) by
continued hydrodynamic escape from the growing planet ({\it Pepin},
1991).  {\it Chen and Ahrens} (1997) estimated that such impacts
produce ground velocities above the escape velocity resulting in the
escape of almost all the atmosphere. However, the question was
revisited by {\it Genda and Abe} (2003): They found that, even in a
collision the size of the Moon-forming impact, less than 30\% of the
atmosphere of both bodies is lost to space. Therefore, giant impacts
can actually result in a net delivery of water to the growing
protoplanet.

\begin{figure}[ht]
\epsscale{1.0}
\plotone{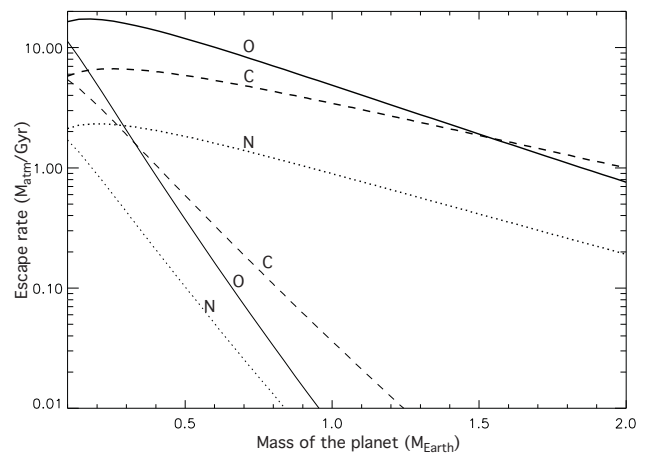}
\caption{\small Mass loss from a terrestrial planet in the habitable
zone of an M starsfor O (solid), N (dotted) and C (dashed) for a
CO$_2$-rich atmosphere with 10\% nitrogen, as a function of the
planetary mass.  The mass loss is given in units of Earth's present
atmosphere per billion years.  Thin lines are for an exosphere
temperature of 10,000~K and thick lines are for 30,000~K.  At these
temperatures, H loss is diffusion limited ({\it Kulikov et al.},
2006). \label{fig.massloss}}
\end{figure}

There may be other, important mechanisms for the removal of volatiles,
including water from the surfaces of otherwise ``habitable'' planets.
The habitable zone of M stars is very close to the star.  Because M
stars tend to have a higher ratio of X-ray and ultraviolet flux to
bolometric flux, radiation and stellar wind-driven escape of planetary
atmospheres may be important.  Exospheric temperatures between 10,000
and 30,000 K are expected. It is within this range of temperature that
Jeans (thermal) escape of the atmosphere is significant.  Figure
\ref{fig.massloss} shows the mass loss from a terrestrial planet for O
(solid), N (dotted) and C (dashed) for a CO$_2$-rich atmosphere with
10\% of nitrogen, as a function of the planetary mass.  (Planets with
high CO$_2$ levels are attractive in this context because the diurnal
temperature difference on the tidally-locked planet is damped.) The
mass loss is given in mass of Earth atmosphere per billion year. Thin
lines are for T$_{exo}$=10,000~K and thick lines for 30,000~K.  At
these temperatures, H loss is of course diffusion-limited.

But around G stars, terrestrial planets may have water abundances much
larger than that of the Earth.  {\it Kuchner} (2003) described another
mechanism of forming water-rich worlds; migration of entire icy
planets inwards by interaction with a gas or planetesimal disk.  Such
``ocean planets'' have also been described by {\it L\'{e}ger et al.}
(2004).  The abundance of water in a planet-forming nebula may have
other secondary, but potentially important implications for
habitability, namely the presence of a giant planet and its dynamical
effects.  For example, the leading theory for the formation of Jupiter
(and some of the habitability properties that it may confer to the
Earth) involves the rapid accretion of a core before depletion of
nebular gas, an accretion accelerated by condensation of water beyond
the ``snow line'' ({\it Stevenson and Lunine}, 1987).  Nebulae with
varying water abundances would presumably be more or less likely to
form gas-accreting cores.

\subsection{\textbf{Planetary composition and diversity}}

\begin{figure}[ht]
\epsscale{1.0}
\plotone{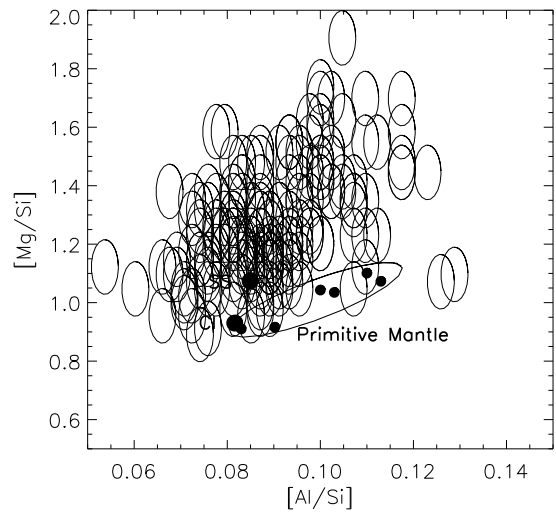}
\caption{\small Plot of Mg/Si vs. Al/Si on hypothetical planet-forming
nebula based on the solar-type star photosphere data of {\it
Edvardsson et al.} (1993).  The circles represent the approximate
range due to measurement errors.  The Edvardsson et al. measurements
are compared with Solar System (SS), chondritic (CI), and several
primitive terrestrial mantle models. Adopted from {\it Gaidos}, in
prep.
\label{fig.silicates}}
\end{figure}

As proposed by {\it Kuchner} (2003) and {\it L\'eger et al.} (2004),
Earth-sized planets around other stars may have very different bulk
compositions than that of our planet.  Even seemingly minor
differences in planetary composition could affect - perhaps
dramatically - geologic activity and geochemical cycles at the
planet's surface.  Just as distance from the Sun, accretion history,
and incorporation of varying amounts of nebular gas have produced a
diversity of planets in our Solar System, we should expect no less
diversity, or probably much more, among a collection of planetary
systems with different cosmochemical inheritances and formation
histories.  For example, two abundant planet-forming elements are
silicon and iron.  Si is an $\alpha$-chain element and produced in
massive stars, whereas Fe is produced primarily in type I SN from
intermediate stars. As a consequence the ratio Fe/Si has increased
with time.  This will influence the size of planetary cores relative
to the mantle as well as the abundance of radiogenic $^{60}$Fe, an
important heat source in the early nebula.  Even the relative
abundances of the major silicate mineral-forming elements (which
controls such properties as melting temperature) vary more from star
to star than they do within the Solar System
(Fig. \ref{fig.silicates}).  Some potential relationships between
cosmochemistry, planetary composition, and habitability have been
discussed by {\it Gonzalez et al.} (2000) and {\it Gaidos et al.}
(2005).  {\it Gaidos} (in prep.) calculated the relative rates of
geologic activity on an Earth whose bulk mantle composition was that
of CI chondrites (perhaps not far from the actual Earth) and a planet
of identical size whose composition was that of enstatite EH chondrite
after the metal has been removed.  The latter has a significantly
higher concentration of the long-lived radioisotopes $^{40}$K,
$^{232}$Th, $^{235}$U, and $^{238}$U ({\it Anders and Grevasse}, 1989;
Newsom, 1995) and such a body would have significantly enhanced rates
of geologic activity, and would remain active for a longer period of
time.

A major parameter that controls the composition of planets is the
ratio of carbon to oxygen (C/O) in the primordial nebula.  Carbon and
oxygen are the two most abundant elements in the interstellar medium
after hydrogen and helium their predominant form in the interstellar
medium is thermodynamically stable CO molecule.  Collapse of molecular
cloud gas leads to higher pressures that favor the formation of water
and methane,
\begin{equation}
{\rm CO + 3H_2 \rightarrow CH_4 + H_2O.}
\end{equation}
However, this reaction is kinetically inhibited on formation time
scales (millions of years) and requires a catalyst such as free iron
({\it Lewis and Prinn}, 1980) If oxygen is more abundant then carbon, then
nearly all C is bound in CO and remaining O is available for the
formation of H$_2$O.  Conversely, excess C results in all O being
bound in CO, absence of H$_2$O, and the formation of graphite and
organic molecules.

The solar photosphere has a C/O of $0.5\pm 0.07$ ({\it Allendo Prieto
et al.}, 2002), and presumably the primordial nebula was oxidizing and
water-rich.  Measurements of C and O abundances in nearby solar-type
stars both with and without planets suggest a significant scatter in
C/O ({\it Gaidos}, in prep.) with the Sun occupying a relatively
C-poor, ``water-rich'' region of the distribution and some stars with
C/O $>$ 1.  Solar-mass stars do not themselves produce significant C
or O, and therefore these abundances reflect that of the gas and dust
(ISM and molecular clouds) from which the stars formed.  Stellar
nucleosynthesis theory predicts that the relative production and
ejection of carbon and oxygen from massive stars (in winds and
supernova ejecta) depends on stellar mass, metallicity, and the amount
of ``dredge-up'' from the carbon-rich interior ({\it Woosley and
Weaver}, 1995).  About 57\% of the C returned to the ISM from a
solar-metallicity stellar population is via the winds of massive
stars: 33\% is produced in intermediate-mass stars and the remainder
in high-mass star ejecta.  Oxygen is almost entirely (87\%) derived
from supernovae and the rest is from their winds.  Molecular clouds
and their offspring can have different C/O because of local
supernovae.  Thus stars and disks that form from the chemically
heterogeneous and evolving interstellar medium will start with
different C/O ratios.  The mean C/O of stellar ejecta increases with
galactic radius such that the older bulge should be more oxygen rich
than the younger disk.  As the Galaxy ages, the C/O ratio of the ISM
and the stars that form from it increases. (Figure \ref{fig.co}).
This picture is consistent with observations of dwarf galaxies ({\it
Garnett et al.}, 1995)

Within a single star-forming region, the C/O can vary because of
condensation and sedimentation of grains ({{\it Lattanzio}, 1984) or
contamination by very massive, short-lived stars within the same
generation.  In fact, the primordial chemistry of the Solar System may
have been influenced by mass loss from nearby massive stars.  {\it
Olive and Schramm} (1982), among others, have suggested that anomalous
Al, Pd, and O isotope ratios in the Solar System can be explained if
the primordial nebula was contaminated with ejected from supernovae,
possibly from short-lived massive stars formed in association with the
Sun.  Local C/O in a planet-forming disk will also be altered by
diffusion of water outwards along the thermal gradient ({\it Cyr et
al.}, 1999) and pile-up of C-rich interstellar dust in the inner
regions of a disk.

\begin{figure}[h]
\epsscale{1.0}
\plotone{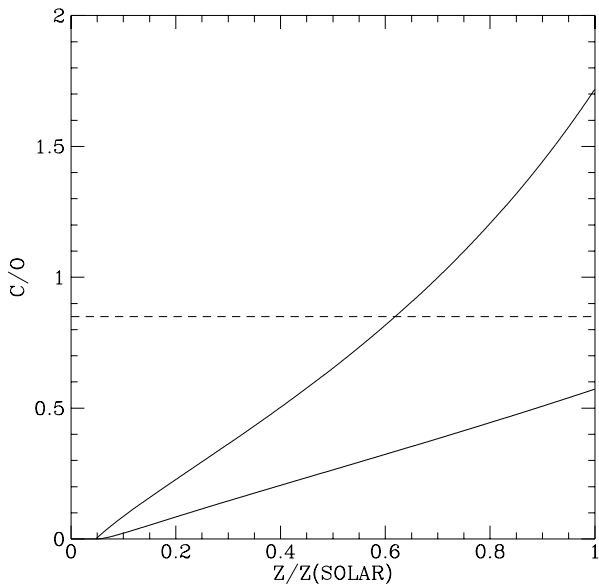}
\caption{\small Calculated evolution of the C/O of stellar wind and
supernova ejecta (top line) and the average interstellar medium
(bottom line) in the disk as a function of the abundance of heavy
elements normalized to the solar value.  The dashed line is the
approximate threshold above which reducing, rather than oxidizing
conditions are expected.  The solar photosphere has C/O of 0.5.
Adopted from Gaidos, in prep.
\label{fig.co}}
\end{figure}

The condensation sequence in a nebula with C/O $\sim 1$ will be
markedly different than that proposed for solar conditions, namely
carbides will replace silicates and carbon will precipitate as
graphite ({\it Larimer}, 1975; {\it Sharp}, 1990).  {\it Gaidos} (2000) suggested that
terrestrial planets would be composed of silicon carbide, a ceramic
with melting temperatures exceeding 3000~K, as well as other carbides.
{\it Seager and Kuchner} (2005) discuss the properties of potential C/O $\gg$
1 planets and calculated an atmospheric spectrum.  They proposed that
the surface of these planets will be covered with organics.  A
``ceramic planet'' will have a Fe-Ni core containing 5-7\% of
dissolved carbon .  Because of the high melting temperature of SiC,
the planet will heat up by a corresponding amount until mantle
convection can remove the heat produced by radiogenic elements.  The
core will be entirely molten and this may mean that such a planet will
lack a magnetic field ({\it Gaidos}, in prep.).  Excess carbon in the mantle will
exist as either graphite, diamond, or liquid carbon, depending on
conditions.  The last will be extremely buoyant and may erupt to the
surface.  Because of the high thermal conductivity of SiC (2-3 times
that of silicates) a thick, rigid lithosphere will develop and plate
tectonics will be less likely.  This example shows that future
searches for other Earths may find instead rather exotic planets.
There is really only one way to find out.

\bigskip

\section{\textbf{EXTRASOLAR EARTHS AND OTHER ORIGINS}}

\subsection{\textbf{Prospects in the Solar System}}

It is difficult to test theories of the origin of life when we are
limited to a single example and when all of the early record of that
life is lost.  Thus searches for a second origin of life outside the
Earth are paramount to understanding our own origins.  Historically,
Mars has been the favorite target in the Solar System; it is the
nearest planet with an accessible surface, and has an atmosphere and
evidence for past geological processes and water.  Initial
disappointment that the Viking missions did not turn up unambiguous
evidence for even simple life forms, and that the surface proved
chemically inhospitable, directed subsequent searches for habitable
conditions (i.e., liquid water) into Mars' past (or most recently with
the MARSIS radar, deep beneath its surface).  Geomorphological
evidence from orbit in the form of outflow channels, valley networks
and possible playa lakes has now been complemented by more direct
geological evidence in the form of aqueous alteration and evaporite
deposition ({\it Squyres et al.}, 2004; {\it Herkenhoff et al.}, 2004;
{\it Klingelh\"{o}fer et al.}, 2004; {\it Rieder et al.}, 2004; {\it
Haskin et al.}, 2005; {\it Hynek}, 2004).  A picture is emerging of a
very early period (of uncertain duration, but perhaps a few hundred
Myr) of a warm, wet Mars, and a cold Mars in the intervening time
({\it Jakosky and Phillips}, 2001; {\it Gaidos and Marion}, 2003; {\it
Solomon et al.}, 2005).  A very exciting possibility is that, due to a
cold climate regime and the absence of plate tectonics, Mars has
preserved information about early prebiotic conditions that has been
lost on Mars.  The oldest planetary rock on Earth is one from Mars
(ALH~84001, 4.5 Ga).

Recent discoveries have also rejuvenated the possibility of habitable
environments on current Mars, albeit at isolated locations in the
subsurface.  These include the presence of abundant regolith ice, the
discovery of ``young'' gully-like formations, and the detection of
atmospheric methane ({\it Mumma et al.}, 2004; {\it Krasnopolsky et
al.}, 2004; {\it Formisano et al.}, 2004).  Methane can be produced
from the high-temperature reduction of CO$_2$ by H$_2$ during
hydrothermal serpentinization of mafic rock ({\it Oze and Sharma},
2005; {\it Lyons et al.}, 2005).  While the possibility of biogenic
methane cannot yet be ruled out, the estimated atmospheric
concentration of a few tens of parts per billion and the lifetime in
the atmosphere ($\sim$300 years) suggest a source flux much weaker
than the estimated abiogenic flux of methane on Earth.  If reports of
latitudinal variation in methane abundance are correct ({\it M. Mumma},
private communication), the lifetime must be much shorter ($\sim$1~yr)
and the flux commensurately higher.  Combined with an upper limit for
SO$_2$ ({\it Krasnopolsky}, 2005) this might disfavor an abiotic seepage
source.  However, Martian geochemistry might be more reducing, thereby
favoring a higher CH$_4$/SO$_2$ ratio, and SO$_2$ disproportionates in
water to sulfate (which is soluble) and hydrogen sulfide (which will
rapidly oxidize to sulfuric acid in the Mars atmosphere).  Regardless,
Mars CH$_4$ gives future astrobiological investigations a focus, e.g.,
measurement of the ratio of stable carbon isotopes to search for
biogenic fractionation.  If life is found on Mars, one possibility is
that it will be unexpectedly familiar.  Efficient ejection and
transfer of material between the planets may have produced a common
ancestry between the planets.  However, if Mars was once habitable and
no evidence for past or present life is found, this constrains models
of lithopanspermia.

Beyond Mars, there are prospects for habitable environments in the
water-rich interiors of the icy satellites of Jupiter, including
Europa and Callisto, and the satellites of Saturn, Titan and
Enceladus.  The debate on the suitability of these objects to support
life centers around the potential energy sources available; while
plausible energy sources are many orders of magnitude lower than the
potential energy from sunlight on Earth ({\it Gaidos et al.}, 1999),
there are several mechanisms by which very low energy fluxes might be
generated in the form of a redox gradient between the atmosphere and
the surface, or between the crust and an interior ocean ({\it Gaidos
et al.}, 1999; {\it Chyba and Phillips}, 2002).  At the minimum, these
bodies offer examples of possible prebiotic chemistries in the Solar
System that might be figuratively and literally frozen in time.
However, the same dynamical barriers, radiation environment, and thick
crust that have isolated these bodies from contamination by
interplanetary transfer of Earth material also challenge the
technologies of humans that choose to investigate these intriguing
environments.

\subsection{\textbf{Extrasolar planets}}

Because the objects in our Solar System are likely to represent a
meager sample of the cosmic diversity of possible habitats for life, a
more complete understanding of the potential abundance and
distribution of life depends on the successful exploration of other
planetary systems.  The Kepler ({\it Borucki et al.}, 2003) and Corot
({\it Bord\'{e} et al.}, 2003) observatories will be capable of
detecting Earth-sized planets as they transit the parent star and will
foreshadow the eventual deployment of far more advanced telescopes
that can directly detect the emitted or reflected light from such
planets.  As spatial resolution of such planets is beyond foreseeable
technology and sources of funding, such characterization will rely on
spectroscopy of their surfaces and atmosphere.  Life manifests itself
by {\it biosignatures}, in this case spectral features of the surface
or atmosphere that reflect its biogeochemical activity and cannot be
found in the absence of life. However, it is possible that abiotic
mechanisms that are not known in the Solar System might reproduce what
was thought to be a reliable biomarker. In fact many features once
claimed to be biosignatures now have convincing abiotic explanations,
e.g. Martian vegetation ({\it Sinton}, 1957) and ``nanobacteria'' in
the ALH 84001 meteorite ({\it McKay et al.}, 1996). The reliability of
a biosignature depends strongly on contextual information.  For
instance the detection of an O$_2$-containing atmosphere does not have
the same implications on the icy moon of a giant planet compared to a
terrestrial planet in the habitable zone of its star ({\it Selsis et
al.}, 2002).  This is because on the latter the weathering of minerals
will consume oxygen and the only source of comparable intensity is
oxygenic photosynthesis.  Conversely, the detection of O$_2$ or O$_3$
is certainly a better biomarker when associated with a reducing
compound such as CH$_4$ or NH$_3$ ({\it Lovelock}, 1975).

\begin{table*}[ht]
\caption{Atmospheric biomarkers: Molecular bands detectable by future
space observatories at infrared [6-20~$\mu$m] and optical
[0.5-0.8~$\mu$m] wavelengths with plausible spectral resolution. A
planet exhibits a biosignature if all the marked bands from a same
line are detected in its spectrum.  Bands in parentheses are
conditional. Empty circles indicate an unlikely but known possible
abiotic origin ({\it Owen}, 1980; {\it L\'eger et al.}, 1993, 1999;
{\it L\'eger}, 2000; {\it Des Marais et al.}, 2002; {\it Selsis et
al.}, 2002; {\it Segura et al.}, 2003; {\it Kasting and Catling},
2003; {\it Selsis et al.}, 2005).}
\label{tab.biomarkers}
\begin{center}
\begin{tabular}{|l|cccccl|ccccl|}
\hline
& \multicolumn{6}{c}{\scriptsize IR  ($\lambda / \Delta \lambda = 25$)} & \multicolumn{5}{|c|}{\scriptsize Visible ($\lambda / \Delta \lambda = 70$)}   \\
\hline & {\scriptsize H$_2$O} & {\scriptsize CO$_2$} & {\scriptsize O$_3$}&{\scriptsize CH$_4$} & {\scriptsize N$_2$O} &{\scriptsize NX} &{\scriptsize H$_2$O} &{\scriptsize O$_2$} & {\scriptsize O$_3$} &{\scriptsize CH$_4$} &  \\
\hline{\scriptsize $\lambda$} & {\scriptsize $<8 $}& {\scriptsize 15} & {\scriptsize 9.6} &{\scriptsize 7.5} & {\scriptsize 7.8} & {\scriptsize ($^{1}$)}  &{\scriptsize 0.72} & {\scriptsize 0.76} & {\scriptsize 0.6} & {\scriptsize 0.73} & \\
{\scriptsize ($\mu$m)}  &{\scriptsize $>18$}	&	&	&	& {\scriptsize 17}	& 	& & {\scriptsize 0.82}		  & {\scriptsize $\pm$0.1} &{\scriptsize 0.79} &\\
\hline{\scriptsize level($^{2}$)}  & {\scriptsize $< 1$} & {\scriptsize $< 1$} & {\scriptsize $< 10$} & {\scriptsize ($^{3}$)} & {\scriptsize $>10$} & {\scriptsize $> 100$}  &{\scriptsize $\leq1$} & {\scriptsize $<10$} & {\scriptsize $\geq 1$} & {\scriptsize $> 50$} & \\
\hline
\hline	
	&\multicolumn{6}{c}{{\scriptsize IR alone}} &  \multicolumn{5}{|c|}{{\scriptsize Visible alone}}  \\
\hline
&$\bullet$ &$\bullet$ &$\bullet$ & & & &{\scriptsize $\circ$}& {\scriptsize $\circ$} & ({\scriptsize $\circ$}) & & {\scriptsize ($^{4}$)} \\
	&$\bullet$	&	&$\bullet$	&$\bullet$	&	&	 &$\bullet$ & $\bullet$	&  ($\bullet$)	&$\bullet$	&  {\scriptsize ($^{4}$)} \\
	&$\bullet$	&	&	&	&$\bullet$	&	&		&	&	&	&	\\
	&$\bullet$	&	&$\bullet$ &	& 	&	&		&	&	&	&	\\
	&$\bullet$	&	&	&$\bullet$	& 	&	&		&	&	&	&	\\
	&{\scriptsize $\circ$}	&	&	&	&	 &{\scriptsize $\circ$}		&	&	&	&	&\\
	\hline
	&\multicolumn{11}{c|}{{\scriptsize Examples of biosignatures requiring both IR and Visible}} \\
	\hline
&({\scriptsize $\circ$})	& {\scriptsize $\circ$}	&	& 	&	   \multicolumn{2}{c}{}      &{\scriptsize $\circ$}	&{\scriptsize $\circ$}	&({\scriptsize $\circ$})	& 	&   {\scriptsize ($^{5}$)}\\
	&$\bullet$	&($\bullet$)	&$\bullet$	&	& 	 \multicolumn{2}{c}{}  &$\bullet$	&	&	&$\bullet$	 &   {\scriptsize ($^{6}$)} \\
 \hline
\multicolumn{12}{l}{{\scriptsize ($^{1}$) NX = NO, NO$_2$ or NH$_3$ - See {\it Selsis et al.} (2005) for wavelengths and required abundance }}  \\	
 \multicolumn{12}{l}{{\scriptsize ($^{2}$) Levels in PAL (Present Atmospheric Level) required for detection at the expected resolution}}  \\
 \multicolumn{12}{l}{{\scriptsize ($^{3}$)1 PAL without H$_2$O, $> 20$~ PAL with H$_2$O}}  \\	
 \multicolumn{12}{l}{{\scriptsize  ($^{4}$) O$_3$ conditional: tracer of O$_2$ - ($^{5}$) Dense CO$_2$ atmosphere:  the IR band of O$_3$ is hidden}} \\
 \multicolumn{12}{l}{{\scriptsize  ($^{6}$) O$_2$ and O$_3$ too low for visible but O$_3$ detected in IR, CH$_4$ hidden by H$_2$O in IR}} \\
\end{tabular}
\end{center}
\label{default}
\end{table*}
Moreover, the absence of a biosignature may not be evidence that a
planet is lifeless, just that a particular metabolism is not present,
that the activity is below detectable limits, or that differences in
the planet's abiotic chemistry mask the biological effect.  Let us
consider that a metabolism $M$ (for instance, oxygenic photosynthesis)
produces a biogenic species $S$ (O$_2$) which, upon accumulation in
the atmosphere can result in a spectral signature $B$ (the 760~nm band
of O$_2$ or the 9.6~$\mu$m of O$_3$). The non-detection of $B$ could
have multiple explanations: 1) Life forms based on $M$ do not exist on
this planet. 2) Life forms based on $M$ do exist but $S$ does not
reach detectable concentrations.  This was probably the case on Earth
between the emergence of O$_2$-producers and the rise of O$_2$, a
period that could have lasted 500-1500 Myr ({\it Catling and Claire},
2005). 3) $S$ reaches levels that would be detectable alone but $B$ is
masked by other spectral features: For instance, the 9.6~$\mu$m O$_3$
band would be masked by the high CO$_2$ level required for greenhouse
warming in most of the habitable zone ({\it Selsis et al.}, 2002b).  The
Lovelock example is another case in point.  The thermodynamic
disequilibrium that Lovelock advocated as a biosignature is a result
of photosynthesis and the conversion of electromagnetic energy into
potential chemical energy in the atmosphere.  In the absence of
photosynthesis, an appropriate biosignature might be the complete {\it
absence} of such a disequilibrium, as this represents an unused source
of energy for microorganisms ({\it Weiss et al.}, 2000).

As a practical approach, one can use the Earth as a reference.  The
spectrum of the Earth exhibits various features that are the direct or
indirect consequence of biological activity. This was demonstrated by
observations from the Galileo probe as it passed near Earth and
detected the presence of O$_2$ (and O$_3$) simultaneously with CH$_4$
({\it Sagan et al.}, 1993).  Table~\ref{tab.biomarkers} gives some
groups of atmospheric molecular bands that could serve as
biosignatures for future missions. Any biomarker should include the
signature of H$_2$O, water being considered as a requisite for life as
we know it.  Some of the listed features are not observable in the
spectrum of present Earth but may have been present in the past. Some
other biogenic compounds, such as N$_2$O were probably never
observable in a low resolution spectrum of the Earth but would be at
slightly higher concentrations.  In addition to atmospheric molecules,
the vegetation ``red edge'' (the increase of plant reflectivity
between 700 and 800~nm) may be another promising way to detect complex
extraterrestrial life ({\it Arnold et al.}, 2002, {\it Seager et al.},
2005). However, the red edge results from photosynthetic pigments like
chlorophyll that are much more complex than simple gases such as
O$_2$.  A life form able to use H$_2$O as an electron donor to reduce
CO$_2$ will produce O$_2$ whatever the pigments or the energy
source. On the other hand, evolution could select other pigments,
characterized by different radiative properties.  Moreover, detecting
the red-edge on a distant Earth replica requires a level of resolution
and sensitivity that will not be reached by the next generation of
telescopes.  There may be other, more readily obtainable pieces of
information contained in the time-variability of emitted or reflected
radiation from a planet about its ability to support life, e.g.,
period of rotation and the presence of an ocean or thick atmosphere
({\it Ford et al.}, 2001; {\it Gaidos and Williams}, 2004; {\it
Williams and Gaidos}, 2005).

Our complete ignorance of when and where life emerged in our Solar
System, and the complexities associated with the maintenance of life
on planetary bodies means that this area of scientific inquiry will be
driven by observations into the foreseeable future.  As a consequence,
the first planet-characterizing missions must be designed for broader
objectives than the search for a specific biomarker.  Perhaps the best
approach is to ``expect the unexpected" and to design instruments not
on the basis of a specific biosignature, but to maximize the potential
for characterization of the physical and chemical properties of the
planet. Inference of biological activity on a planet could emerge from
a more general understanding of its spectrum, even if none of the
expected biosignatures are found.  For the foreseeable future, a
working definition for the biosignatures of remote, inaccessible
planetary life might remain {\it chemical phenomena that cannot be
explained by all known abiotic chemistry}.  This is ultimately an
unsatisfactory state of affairs but we should not despair too quickly:
Not quite four centuries have elapsed since Galileo turned his
telescope to the other planets in our Solar System and it has been a
mere decade since the discovery of the first extrasolar planet around
a main-sequence star.  Should our species choose to desist from
threatening the life and habitability of this world, our progeny will
have the fullness of time to answer the question of whether other
planets host living beings and whether any of them also ponder the
same question.

\bigskip

\textbf{ Acknowledgments.} EG thanks the Centre Recherche Astronomique
de Lyon, the NASA Terrestrial Planet Finder Foundation Science
program, and the NASA Astrobiology Institute for travel support during
this chapter's writing. Both authors thank Bruce Jakosky for a helpful
review.

\bigskip

\centerline\textbf{REFERENCES}
\bigskip
\parskip=0pt
{\small
\baselineskip=11pt

\refs Adams F.~C., and Spergel D.~N. (2005) {\it Astrobiol., 5},
497-514.

\refs Alexander C.~M.~O'D., Russell S.~S., Arden J.~W., Ash R.~D.,
Grady M.~M., and Pillinger C.~T. (1998) {\it Meteor. Planet. Sci.,
33}, 603-622.

\refs Allende-Prieto C., Lambert D.~L., and Asplund M. (2002) {\it
Astrophys. J., 573}, L137-L140.

\refs Anbar A.~D., Zahnle K.~J., Arnold G.~L., and Mojzsis
S.~J. (2001) {\it J. Geophys. Res., 106}, 3219-3236.

\refs Anders E., DuFresne A., Fitch F.~W., Cavaill\'{e} A.,
Dufresne E.~R., and Hayatsu R. (1964) {\it Science, 146}, 1157-1161.

\refs Armitage P.~J., Clarke C.~J., and Palla F. (2003) {\it
Mon. Not. R. Astron. Soc., 342}, 1139-1146.

\refs Arnol, L., Gillet S., Lardi\'ere O., Riaud P., and Schneider,
J. (2002) {\it Astron. Astrophys., 392}, 231-237.

\refs Asghari N., Broeg C., Carone L., Casa-Miranda R., Castro Palacio
J.~C., et al. (2004) {\it Astron. Astrophys., 426}, 353-365.

\refs Bada J.~L., Bigham C., and Miller S.~L. (1994) {\it
Proc. Natl. Acad. Sci. USA, 91}, 1248-1250.

\refs Belbruno E., and Gott III J.~R. (2005) {\it Astron. J., 129},
1724-1745.

\refs Bhattacharya J.~P., Payenberg T.~H.~D., Lang S.~C., and
Bourke M. (2005) {\it Geophys. Res. Lett., 32}, L10201.

\refs Bord\'{e} P., Rouan D., and L\'{e}ger A. (2003) {\it
Astron. Astrophys., 405}, 1137-1144.

\refs Borucki W.~J., Koch D.~G., Lissauer J.~J., Basri G.~B., Caldwell
J.~F., Cochran W.~D., Dunham E.~W., Geary J.~C., Latham D.~W.,
Gilliland R.~L., Caldwell D.~A., Jenkins J.~M., and Kondo Y. (2003)
{\it Proc. SPIE Conf., 4854}, 129-140.

\refs Bowring S.~A., and Williams I.~S. (1999) {\it
Contrib. Mineral. Petrol., 134}, 3-16.

\refs Boyet M., and Carlson R.~W. (2005) {\it Science, 309}, 576-581.

\refs Brasier M., Green O., Lindsay J., and Steele A. (2004) {\it
Orig. Life Evol. Biosph., 34}, 257-269.

\refs Brasier M.~D., Green O.~R., Jephcoat A.~P., Kleppe A.~K.,
van Krenedonk M.~J., Lindsay J.~F., Steel A., and Grassineau
N.~V. (2002) {\it Nature, 416}, 76-81.

\refs Brochier C., and Philippe H. (2002) {\it Nature, 417}, 244.

\refs Burchell M.~J., Shrine N.~R.~G., Mann J., Bunch A.~W.,
Brand~{a}o P., Zarneck J.~C., and Galloway J.~A. (2001) {\it
Adv. Space Res., 28}, 707-712.

\refs Burchell M.~J., Galloway J.~A., Bunch A.~W., and Brand\~{a}o
P.~F.~B. (2003) {\it Orig. Life Evol. Biosph., 33}, 53-74.

\refs Burchell M.~J., Mann J.~R., and Bunch A.~W. (2004) {\it
Mon. Not. R. Astron. Soc., 352}, 1273-1278.

\refs Caldeira K., and Kasting J.~F. (1992) {\it Nature, 360}, 721-723.

\refs Carr M.~H. (1999) {\it J. Geophys. Res., 104}, 21897-21910.

\refs Castresana J., L\"{u}bben M., Saraste M., and Higgins
D.~G. (1994) {\it EMBO J., 13}, 2516-2525.

\refs Catling D.~C. (2006) {\it Science, 311}, 38a.

\refs Catling D.~C., and Claire M.~W. (2005) {\it Earth
Planet. Sci. Lett., 237}, 1-2.

\refs Cech T.~R. (1986) {\it Cell, 44}, 207-210.

\refs Chambers J.~E. (2001) {\it Icarus, 152}, 205-224

\refs Chambers J.~E., and Wetherill G.~W. (1998) {\it Icarus, 136}, 304-327.

\refs Chen G.~Q., and Ahrens T.~J. (1997) {\it Phys. Earth
Planet. Int., 100}, 21-26.

\refs Chyba C.~F., and Hand K.~P. (2005) {\it
Ann. Rev. Astron. Astrophys., 43}, 31-74.

\refs Chyba C.~F., and Phillips C.~B. (2002) {\it Origins Life
Evol. Biosph., 32}, 47-67.

\refs Chyba C.~F., Whitmire D.~P., and Reynolds R. (2000) In {\it
Protostars and Planets IV}, (V. Mannings, A.~P. Boss, and
S.~S. Russell, eds.), pp. 1365-1393, Univ. Arizona, Tucson.

\refs Chyba C.~F., Thomas P.~J., Brookshaw L., and Sagan C. (1990)
{\it Science, 249}, 366-373.

\refs Clark B.~C., Baker A.~L., Cheng A.~F., Clement S.~J., McKay D.,
McSween H.~Y., Pieters C.~M., Thomas P., and Zolensky M. (1999) {\it
Orig. Life Evol. Biosph., 29}, 521-545.

\refs Cody G.~D. (2004) {\it Ann. Rev. Earth Planet. Sci., 32}, 569-599.

\refs Cody G.~D., Alexander C.~M.~O., and Tera F. (2002) {\it
Geochim. Cosmochim. Acta, 66}, 1851-1865.

\refs Cody G.~D., Boctor N., Filley T.~R., Hazen R.~M., Scott
J.~H., Sharma A., and Yoder H.~S., Jr. (2001) {\it Science, 289},
1337-1340.

\refs Cohen B.~A., Swindle T.~D., and Kring D.~A. (2000) {\it Science,
290}, 1754-1755.

\refs Commeyras A., Taillades J., Collet H., Boiteau L.,
Vandenbeele-Trambouze O., et al. (2004) {\it Orig. Life
Evol. Biosph., 34}, 35-55.

\refs Cooper G., Kimmich N., Bellsle W., Sarinana J., Brabham, K., and
Garrel, L.(2001) {\it Nature, 414}, 879-883.

\refs Corliss J.~B., Dynmond J., Gordon L.~I., Edmont J.~M., von
Herzen R.~P., Ballard R.~D., Green K., Williams D., Bainbridge
A., Crane K., and van Andel T.~H. (1979) {\it Science, 203}, 1073-1083.

\refs Craddock R.~A. and Howard A.~D. (2002) {\it
J. Geophys. Res., 107}, 5111.

\refs Crick F.~H.~C. (1968) {\it J. Mol. Biol., 38}, 367-379.  

\refs Cuntz M. (2002) {\it Astrophys. J., 572}, 1024-1030.

\refs Cuzzi J.~N. and Zahnle K.~J. (2004) {\it Astrophys. J., 614},
490-496.

\refs Cyr K.~E., Sharp C.~M., and Lunine, J.~I. (1999) {\it
J. Geophys. Res., 104}, 19,003-19,014.

\refs Cyr K.~E., Sears W. D., and Lunine J.~I. (1998) {\it Icarus,
135}, 537-548.

\refs Dalrymple G.~B. and Ryder G. (1993) {\it J. Geophys. Res., 98},
13,085-13,095.

\refs Dauphas N., Robert F., and Marty B. (2000) {\it Icarus, 148},
508-512.

\refs Davies P.~C.~W. and C.~H. Lineweaver (2005) {\it Astrobiol., 5},
154-163.

\refs Des Marais D.~J., Harwit M.~O., Jucks K.~W., Kasting J.~F.,
Lin D.~N.~C., Lunine J.~I., Schneider J., Seager S., Traub W.~A.,
and Woolf N.~J. (2002) {\it Astrobiol., 2}, 153-181.

\refs Di Giulio M. (2003) {\it J. Mol. Evol., 57}, 721-730.

\refs Ecuvillon A., Israelian G., Santos N.~C., Mayor M., Villar
V., and Bihain G. (2004) {\it Astron. Astrophys., 426}, 6619-6630.

\refs Ecuvillon A., Israelian G., Santos N.~C., Shckukina N.~G., Mayor
M., and Rebolo R. (2005) {\it Astron. Astrophys.}, in press.

\refs Edvarsson B., Andersen J., Gustafsson B., Lambert D.~L., Nissen
P.~E., and Tomkin J. (1993) {\it Astron. Astrophys., 275}, 101-152

\refs Engel M.~H. and Nagy B. (1982) {\it Nature, 296}, 837-840.

\refs \'{E}rdi B., Dvorak R., S\'{a}ndor Z., Pilat-Lohinger E.,
and Funk B. (2004) {\it Mon. Not. R. Astron. Soc., 351}, 1043-1048.

\refs Fedo C.~M. and Whitehouse M.~J. (2002) {\it Science, 296},
1448-1452.

\refs Fernandez J.~A. and Ip W.~H. (1983) {\it Icarus, 54}, 377

\refs Ford E.~B., Seager S., and Turner E.~L. (2001) {\it Nature,
412}, 885-887.

\refs Formisano V., Atreya S., Encrenaz T., Ignatiev N., and
Giuranna M. (2004) {\it Science, 306}, 1758-1761.

\refs Freeland S. J., Wu T., and Keulmann N. (2003) {\it Orig. Life
Evol. Biosph., 33}, 457-477.

\refs Furnes H., Banerjee N.~R., Muehlenbachs K., Staudigel H., and de
Wit, M. (2004) {\it Science, 304}, 578-581.

\refs Gaidos E.~J. (2000) {\it Icarus, 145}, 637-640.

\refs Gaidos E.~J., and Marion, G. (2003) {\it J. Geophys. Res., 108},
DOI 10.1029/2002JE002000.

\refs Gaidos E. and Williams D.~M. (2004) {\it New Astron., 10}, 67-77.

\refs Gaidos E. , Deschenes B., Dundon L., Fagan K., McNaughton C.,
Menviel-Hessler L., Moskovitz N., and Workman M. (2005) {\it
Astrobiol., 5}, 100-126.

\refs Galtier N., Tourasse N., and Gouy M. (1999) {\it Science, 283},
220-221.

\refs Garnett D.~R., Skillman E.~D., Dufour R.~J., Peimbert M.,
Torres-Peimbert S., Terlevich R., Terlevich E., and Shields
G. A. (1995) {\it Astrophys. J., 443}, 64-76.

\refs Gehman C.~S., Adams F.~C., and Laughlin G. (1996) {\it
Publ. Astron. Soc. Pac., 108}, 1018-1023.

\refs Genda H. and Abe Y. (2003) {\it Icarus, 165}, 149-162.

\refs Gilbert W. (1986) {\it Nature, 319}, 618.

\refs Gladman B., Dones L., Levison H.~F., and Burns J.~A. (2005)
{\it Astrobiol., 5}, 483-496.

\refs Gladman B., and Burns J.~A. (1996) {\it Science, 274}, 161-162.

\refs Goldreich P., Lithwick Y., and Sari R. (2004) {\it
Astrophys. J., 614}, 497-507.

\refs Gonzalez G., Brownlee D., and Ward P. (2001) {\it Icarus, 152},
185-200.

\refs Goswami J.~N., Sinha N., Murty S.~V.~S., Mohapatra R.~K.,
and Clement C.~J. (1997) {\it Meteorit. Planet. Sci., 32}, 91-96.

\refs Gozdziewski K. (2002) {\it Astron. Astrophys., 393}, 997-1013.

\refs Grotzinger J.~P., and Rothman D.~H. (1996) {\it Nature, 383},
423-425.

\refs Hahn J.~M. and Malhotra R. (1999) {\it Astron. J., 117},
3041-3053.

\refs Hart M.~H. (1979) {\it Icarus, 37}, 351-357.

\refs Haskin L.~A., Wang A., Jolliff H.~Y., McSween H.~Y., Clark,
D.~J., et al. (2005) {\it Nature, 436}, 66-69.

\refs Herkenhoff K.~E., Squyres S.~W., Arvidson R., Bass D.~S., Bell
III J.~F., et al. (2004) {\it Science, 306}, 1727-1730.

\refs Holm N.~G. and Andersson E. (2005) {\it Astrobiol. 5}, 444-460.

\refs Holzheid A., Sylvester P., O'Neill H.~S.~C., Rubie D.~C., and
Palme H. (2000) {\it Nature, 406}, 396-399.

\refs Horneck G., Rettberg P., Reitz G., Wehner J., Eschweiler
U., Strauch K., Panitz C., Starke V., and Baumstark-Khan C. (2001)
{\it Orig. Life Evol. Biosph., 31}, 527-546.

\refs Huang X., Xu Y., and Karato S.-I. (2005) {\it Nature, 434},
746-749.

\refs Huang S.-S. (1959) {\it Am. Sci., 47}, 392-402.

\refs Huber C. and W\"{a}chtersh\"{a}user G. (1997) {\it Science,
276}, 245-247.

\refs Hynek B.~M. (2004) {\it Nature, 431}, 156-159

\refs Jacobsen S.~B. (2005) {\it Ann. Rev. Earth Planet. Sci., 33},
531-570.

\refs Jakosky B.~M. and Phillips, R. (2001) {\it Nature, 412},
237-244.

\refs Ji J., Liu L., Kinoshita H., and Li G. (2005) {\it
Astrophys. J., 631}, 1191-1197.

\refs Johnson D.~C., Dean D.~R., Smith A.~D., and Johnson M.~K. (2005)
{\it Ann. Rev. Biochem., 74}, 247-281.

\refs Jones B.~W., Underwood D.~R., and Sleep P.~N. (2005) {\it
Astrophys. J., 622}, 1091-1101.

\refs Jones B.~W. and Sleep P.~N. (2002) {\it Astron. Astrophys.,
393}, 1015-1026.

\refs Joshi M.~M., Haberle R.~M., and Reynolds R.~T. (1997) {\it
Icarus, 129}, 450-465.

\refs Joyce G. (2004) {\it Ann. Rev. Biochem., 73}, 791-836.

\refs Kasting J.~F. and Catling D. (2003) {\it
Ann. Rev. Astron. Astrophys., 41}, 429-463.

\refs Kasting J.~F., Whitmire D.~P., and Reynolds R.~T. (1993) {\it
Icarus, 101}, 108-128.

\refs Kegler P., Holzheid A., Rubie D.~C., Frost D., and Palme H.
(2005) {\it Lunar Planet. Sci. Conf. Abst., 36} 2030.

\refs Klingelh\"{o}fer G., Morris R.~V., Bernhardt B., Schr\"{o}der
C., Rodionov, D.~S., et al. (2004) {\it Science, 306}, 1740-1745.

\refs Krasnopolsky, V~A. (2005) {\it Icarus, 178}, 487-492.

\refs Krasnopolsky V.~A., Maillard J.~P., and Owen T.~C. (2004) {\it
Icarus, 172}, 537-547.

\refs Kuchner M.~J. (2003) {\it Astrophys. J., 596}, L105-L108.

\refs Kulikov Y.~N., Lammer H., Lichtenegger H.~I.~M., Terada N.,
Ribas I., Kolb C., Langmayr D., Lundin R., Guinan E.~F.,
Barabash S., and Biernat H.~K. (2006) {\it Icarus}, in press.

\refs Kvenvolden K., Lawless J., Pering K., Peterson E., Flores J.,
Ponnamperuma C., Kaplan I.~R., Moore C. {\it Nature, 228}, 923-926.

\refs Larimer J.~W. (1975) {\it Geochim. Cosmochim. Acta, 39},
389-392.

\refs Laskar J. (1994) {\it Astron. Astrophys., 287}, L9-L12.

\refs Lathe R. (2004) {\it Icarus, 168}, 18-22.

\refs Lattanzio J.~C. (1984) {\it Mon. Not. R. Astron. Soc., 207},
309-322.

\refs Laughlin G., Chambers J., and Fischer D. (2002) {\it
Astrophys. J., 579}, 455-467.

\refs L\'{e}cuyer C.(1998) {\it Chem. Geol., 145}, 249-261.

\refs L\'{e}ger A., Selsis F., Sotin C., Guillot T., Despois D.,
Mawet D., Ollivier M., Lab\`{e}que A., Valette C., Brachet F.,
Chazelas B., and Lammer H. (2004) {\it Icarus, 169}, 499-504.

\refs L\'{e}ger A. (2000) {\it Adv. Space Res. 25}, 2209-2223.

\refs L\'{e}ger A., Ollivier M., Altwegg K., and Woolf
N.~J. (1999) {\it Astron. Astrophys., 341}, 304-311.

\refs L\'{e}ger A., Pirre M., and Marceau F.~J. (1993) {\it
Astron. Astrophys., 277}, 309-313.

\refs Levy M. and Miller S.~L. (1998) {\it Proc. Natl. Acad. Sci. USA,
95}, 7933-7938.

\refs Lewis J.~S. and Prinn, R.~G. (1980) {\it Astrophys. J., 238},
357-364.

\refs Lineweaver C.~H. and Davis T. (2002) {\it Astrobiol., 2},
293-304.

\refs Litasov K., Ohtani E., Langenhorst F., Yurimoto H., Kubo T., and
Kondo T. (2003) {\it Earth Planet. Sci. Lett., 211}, 189-203.

\refs Lodders K. (2004) {\it Astrophys. J., 611}, 587-597.

\refs Lovelock J.~E. (1975) {\it Proc. R. Soc. Lond. B, 189}, 167-.

\refs Lunine J., Chambers J., Morbidelli A., and Leshin L. (2003) {\it
Icarus, 165}, 1-8.

\refs Lyo A.-R., Lawson W.~A., Mamajek E.~E., Feigelson E.~D., Sung
E.-C., and Crause L.~A. (2003) {\it Mon. Not. R.  Astron. Soc., 338},
616-622.

\refs Lyons J.~R., Manning C., and Nimmo F. (2005) {\it
Geophys. Res. Lett., 32}, L13201.

\refs Maher K.~A. and Stevenson D.~J. (1988) {\it Nature, 331},
612-614.

\refs Marchis F., Descamps P., Hestroffer D., Berthier J., and de
Pater I. (2005) {\it Bull Amer. Astron. Soc. DPS, 36}, 46.02

\refs Mastrapa R.~M.~E., Glanzberg H., Head J.~N., Melosh H.~J., and
Nicholson W.~L. (2001) {\it Earth Planet Sci. Lett., 189}, 1-8.

\refs Matsui T. and Abe Y. (1986) {\it Nature, 322}, 526-528.

\refs McFadden G.~I. (2001) {\it J. Phycol., 37}, 951-959.

\refs McKay D.~S., Gibson, Jr. E.~K., Thomas-Keprta K.~L., Hojatollah
V., Romanek C.~S., Clemett S.~J., Chillier X.~D.~F., Maechling C.~R.,
and Zare R.~N. (1996) {\it Science, 273}, 924-930.

\refs Melosh H.~J. (2003) {\it Astrobiol., 3}, 207-215.

\refs Melosh H.~J. (1984) {\it Icarus, 59}, 234-260.

\refs Menou K. and Tabachnik S. (2003) {\it Astrophys. J., 583},
473-488.

\refs Miller S.~L. (1953) {\it Science, 117}, 528-529.

\refs Miller S.~L. and Schlesinger G. (1983) {\it Adv. Space Res., 3},
47-53.

\refs Miyakawa S., Cleaves H.~J., and Miller S.\~L. (2002a) {\it
Orig. Life Evol. Biosph., 32}, 195-208.

\refs Miyakawa S., Cleaves H.~J., and Miller S.~L. (2002b) {\it
Orig. Life Evol. Biosph., 32}, 209-218.

\refs Mojzsis S.~J., Arrenhius G., McKeegan K.~D., Harrison T.~M.,
Nutman A.~P., and Friend C.~R.~L. (1996) {\it Nature, 384}, 55-59.

\refs Mojzsis S.~J., Harrison T.~M., Friend C.~R.~L., Nutman A.~P.,
Bennett V.~C., Fedo C.~M., and Whitehouse M.~J. (2002) {\it Science,
298}, 917a.

\refs Morbidelli A., Chambers J., Lunine J.~I., Petit J.~M., Robert
F., Valsecchi G.~B., and Cyr K.~E. (2000) {\it Meteorit. Planet. Sci.,
35}, 1309-1320.

\refs Mousis O. and Alibert Y. (2005) {\it Mon. Not. R.  Astron. Soc.,
358}, 188-192.

\refs Mumma M.~J., Novak R.~E., DiSanti M.~A., Bonev B.~P., Dello
Russo N. (2004) {\it Bull. Amer. Astron. Soc. DPS, 36} 26.02

\refs Newsom H.~E. (1995) in {\it Global Earth Physics: A Handbook of
Physical Constants}, (T. J. Ahrends, ed.), pp. 159-189, AGU,
Washington, DC.

\refs Nisbet E.~G. and Sleep N.~H. (2001) {\it Nature, 409},
1083-1091.

\refs Olive K.~A. and Schramm D.~N. (1982) {\it Astrophys. J., 257},
276-282.

\refs Orgel L.~E. (2004) {\it Orig. Life Evol. Biosph., 34}, 361-369.

\refs Orgel L.~E. (1968) {\it J. Mol. Biol., 38}, 381-393.

\refs Owen T. (1980) In {\it Strategies for the search for life in
the Universe} (M. Papagiannis, ed.), pp. 177-188, Reidel, Dordrecht.

\refs Oze C. and Sharama M. (2005) {\it Geophys. Res. Lett., 32},
L10203.

\refs Pahlevan K. and Stevenson D. J. (2005) {\it Lunar Planet.
Sci. Conf. DPS, 36}, Abst. 2382.

\refs Pepin R. O. (1991) {\it Icarus, 92}, 2-79.

\refs Podesek F.~A. and Ozima M. (2000) In {\it Origin of the Earth
and Moon}, (R. M. Canup and K. Righter, eds.), pp. 63-72,
Univ. Arizona, Tucson.

\refs Petit J.-M., Morbidelli A., and Chambers J. (2001) {\it Icarus,
153}, 338-347.

\refs Raymond S.~N., Quinn T., and Lunine J.~I. (2004) {\it Icarus,
168}, 1-17.

\refs Raymond S.~N., Quinn T., and Lunine J.~I. (2005) {\it
Astrophys. J., 632}, 670-676.

\refs Ribas I., Guinan E.~F., G\"{u}del M., and Audard M. (2005) {\it
Astrophys. J., 622}, 680-694.

\refs Rieder R., Gellert R., Anderson R.~C., Br\"{u}ckner J., Clark
B.~C., et al. (2004) {\it Science, 306}, 1746-1749.

\refs Righter K. (2003) {\it Ann. Rev. Earth Planet. Sci., 31},
135-174.

\refs Rosing M.~T. (1999) {\it Science, 283}, 674-676.

\refs Russell M.~J. and Arndt N.~T. (2005) {\it Biogeosci., 2},
97-111.

\refs Sagan C., Thompson W.~R., Carlson R., Gurnett D., and Hord C.
(1993) {\it Nature, 365}, 715-721.

\refs Schidlowski M.~A. (1988) {\it Nature, 333}, 313-318.

\refs Schaefer L. and Fegley, Jr. B. (2005) {\it Bull
Amer. Astron. Soc. DPS, 37}, 29.15

\refs Schopf J.~W. and Packer B.~M. (1987) {\it Science, 237}, 70-73.

\refs Schulte M.~D. and Shock E.~L. (1992) {\it Meteorit., 27}, 286.

\refs Seager S., Turner E.~L., Schafer J. and Ford E.~B. (2005) {\it
Astrobiol., 5}, 372-390.

\refs Segura A., Krelove K., Kasting J.~F., Sommerlatt D., Meadows V.,
Crisp D., Cohen M., and Mlawer E. (2003) {\it Astrobiol., 3}, 689-708.

\refs Selsis F., Despois D. and Parisot, J.-P. (2002) {\it
Astron. Astrophys., 388}, 985-1003.

\refs Selsis F. (2002) {\it ASP Conf. Ser., 269}, 273-281

\refs Sharp C. M. (1990) {\it Astrophys. Space Sci., 171}, 185-188.

\refs Shock E.~L. and Schulte M.~D. (1998) {\it J. Geophys., 103},
28513-28528.

\refs Shock E.~L. and Shulte M.~D. (1990) {\it Nature, 343}, 728-731.

\refs Shock E.~L., Amend J.~P., and Zolotov M.~Y. (2000) In {\it The
Origin of the Earth and Moon}, (R. Canup and K. Righter, eds.),
pp. 527-543, Univ. Arizona, Tucson.

\refs Sinton W.~. (1957) {\it Astrophys. J., 126}, 231-239.

\refs Sleep N.~H., Meibom A., Fridriksson Th., Coleman R.~G. and Bird
D.~K. (2004) {\it Proc. Natl. Acad. Sci. USA, 101}. 12,818-12,823.

\refs Sleep N.~H. and Zahnle K. (1998) {\it J. Geophys. Res., 103},
28529-28544.

\refs Solomon S.~C., Aharonson O., Aurnou J.~M., Banerdt W.~B., Carr
M.~H., et al. (2005) {\it Science, 307}, 1214-1220.

\refs Squyres S.~W., Arvidson R.~E., Bell, III J.~F., Br\"{u}ckner J.,
Cabrol N.~A., et al. (2004) {\it Science, 306}, 1709-1714.

\refs Stevenson D.~J. (1980) {\it Lunar Planet. Sci. Conf., 11},
1088-1090.

\refs Strom R.~G., Malhotra R., Ito T., Yoshida F., and Kring
D.~A. (2005) {\it Science, 309}, 1847-1850.

\refs Tarits P., Hautot P. and Perrier F. (2004) {\it
Geophys. Res. Lett., 31}, L06612.

\refs Taylor S.~R. (1999) {\it Meteor. Planet. Sci., 34}, 317-239.

\refs Tian F., Toon O.~B., Pavlov A.~A., and De Sterck H. (2005) {\it
Science, 538}, 1014-1017.

\refs Tolstikhin I.~N. and O'Nions R.~K. (1994) {\it Chem. Geol.,
115}, 1-6.

\refs Travis B. J. and Schubert G. (2005) {\it Earth
Planet. Sci. Lett., 240}, 234-250.

\refs Tsiganis K., Gomes R., Morbidelli A., and Levison H.~G. (2005)
{\it Nature, 435}, 459-461.

\refs Turner G., Knott S.~F., Ash R.~D., and Gilmour J.~D. (1997) {\it
Geochim. Cosmochim. Acta, 61}, 3835-3850.

\refs van Zuilen M.~A., Lepland A. and Arrhenius G. (2002) {\it
Nature, 418}, 627-630.

\refs Vlassov A.~V., Kazakov S.~A., Johnston B.~H. and Landweber
L.~F. (2005) {\it J. Mol. Evol., 61}, 264-273.

\refs Walker J.~C.~G., Hays P.~B., and Kasting J.~F. (1981) {\it
J. Geophys. Res., 86}, 9776-9782.

\refs Wallis M.~K. and Wickramasinghe N.~C. (2004) {\it
Mon. Not. R. Astron. Soc., 348}, 52-61.

\refs Weidenschilling S.~J. (1977) {\it Astrophys. Space Sci., 51},
153-158.

\refs Weidenschilling S.~J. (1981) {\it Icarus, 46}, 124-126.

\refs Weiss B.~P., Kirschvink J.~L., Baudenbacher F.~J., Vali H.,
Peters N., Macdonald F.~A., and Wikswo J.~P (2000) {\it Science, 290},
791-795.

\refs Weiss B.~P., Yung Y.~L. and Nealson K.~H. (2000) {\it
Proc. Natl. Acad. Sci. USA, 97}, 1395-1399.

\refs Wells L.~E., Armstrong J.~C., and Gonzalez G. (2003) {\it
Icarus, 162}, 38-46.

\refs Wiechert U., Halliday A.~N., Lee D.-C., Snyder G.~A., Taylor
L.~A., and Rumble D. (2001) {\it Science, 294}, 345-348.

\refs Wilde S.~A., Valley J.~W., Peck W.~H., and Graham C.~M. (2001)
{\it Nature, 409}, 175-178.

\refs Williams D.~M. and Gaidos E. (2005) {\it
Bull. Amer. Astron. Soc. DPS, 37}, 31.13.

\refs Williams D.~M. and Pollard D. (2003) {\it Int. J. Astrobiol.,
2}, 1-19.

\refs Williams D.~M., and Pollard D. (2002) {\it Int. J. Astrobiol.,
1}, 61-69.

\refs Woese C.~R. (2000) {\it Proc. Natl. Acad. Sci. USA, 97},
8392-8396.

\refs Woese C.~R. (1987) {\it Microbiol. Rev., 51}, 221-271.

\refs Woosley S.~E. and Weaver T.~A. (1995) {\it Astrophys. J. Supp.,
101}, 181-234.

\refs Yarus M., Caporaso J.~G., and Knight R. (2005) {\it
Ann. Rev. Biochem., 74}, 179-198.

\refs Zahle K. (1998) in {\it Origins, ASPC 148} (C. E. Woodward,
J. M. Shull, and H. A. Thronson, Jr., eds.), pp. 364-391, ASP, San
Francisco.

\refs Zolotov M.~Y. and Shock E.~L. (2000) {\it Icarus, 150}, 323-337.

\end{document}